\documentclass[12 pt]{article}
\usepackage[utf8x]{inputenc}
\usepackage{amsmath}
\usepackage{amsfonts}
\usepackage{amssymb}
\usepackage{graphicx}
\usepackage{empheq}
\usepackage[sort&compress,numbers]{natbib}
\usepackage{doi}
\hoffset= - 0.9in         

\newcommand{\p}{\partial}
\newcommand{\f}{\frac}
\newcommand{\s}{\sqrt}

\newcommand{\be}{\beta}
\newcommand{\al}{\alpha}

\newcommand{\cf}{\mathcal{F}}

\newcommand{\cm}{\mathcal{M}}

\newcommand{\g}{\Gamma}

\newcommand{\nn}{\nonumber}

\newcommand{\ket}[1]{\left| #1 \right>} 

\usepackage{hyperref}
\usepackage{tikz}
\usepackage{epigraph}
\usepackage{float}
\begin{document}
\title{On modular transformations of non-degenerate toric conformal blocks}
 \author{{ N.Nemkov}\thanks{{\small {\it Moscow Institute of Physics and Technology (MIPT), Dolgoprudny, Russia}
 and {\it Institute for Theoretical and Experimental Physics (ITEP), Moscow, Russia}; nnemkov@gmail.com} }
\date{ }}
\date{\today}
\maketitle
\vspace{-5.0cm}

\begin{center}
 \hfill ITEP/TH-8/15\\
\end{center}

\vspace{3.5cm}
\begin{abstract}
We derive and solve the difference equations on the toric modular kernel following from the consistency relations in the fusion algebra. The result is explicit and simple series expansion for the toric modular kernel of non-degenerate Virasoro conformal blocks. We show that this expansion is equivalent to the celebrated integral representation due to B. Ponsot and J. Teschner. We also interpret obtained series representation as a non-perturbative expansion and note that this raises further questions.
 
\end{abstract}
\newpage
 \tableofcontents
\newpage

\section{Introduction}
Conformal blocks are important ingredients for a wide variety of physical theories. Though being fully determined by the conformal symmetry they are not completely understood. A number of representations allowing for control over some of their properties, together with a limited amount of the exactly solvable examples exists. In this paper we study a particular property of the conformal blocks, namely, their behavior under the modular transformations. This direction seems especially interesting for  the following reason. First, usual definition of the conformal blocks keeps their properties under the modular transformations well hidden. Second, modular transformations can be studied indirectly, without calling for the explicit shapes of conformal blocks. Therefore, by investigating the modular transformations of conformal blocks we can hope to gain a deeper insight at the aspects of their structure which are not at all visible from the definition or conventional representations.

It was shown recently \cite{GMMpert, Billo:2013fi, Lerda, Nemkov1} that \textit{perturbatively} modular transformations of conformal blocks are nothing but the Fourier transformation. In the present paper we construct a \textit{non-perturbative} expansion of the modular kernel on top of this asymptotic Fourier form and show, that the resulting series is equivalent to the integral representation suggested in papers \cite{PT1,PT2,PT3}. As mentioned, the method to obtain this non-perturbative series is indirect. To find the origin of these non-perturbative corrections in the conformal blocks themselves remains an open question.  

Organization of the paper is as follows. In section \ref{sec conformal blocks} we remind the definition of conformal blocks and elaborate on the existing exact examples. In section \eqref{sec modular transformations} we remind basic facts about the modular transformations and give illustrations for the exact conformal blocks. In section \eqref{sec PT formula} we describe the integral formula from papers \cite{PT1, PT2, PT3} for the modular transformations of the \textit{generic} (i.e. not necessarily degenerate) conformal blocks. In section \ref{sec main result} we present our main result which can be viewed as a series representation of the integral formula. In subsequent section \ref{sec non-perturbative discussion} we discuss the series representation as a non-perturbative expansion. In sections \ref{sec consistency constraints}, \ref{sec series from integral} we derive the series representation in two independent ways. First, by solving consistency constraints on the modular kernel and second, by explicitly evaluating the integral of \cite{PT1,PT2,PT3}. Appendix \ref{app special functions} presents special functions we use in the main text while appendix \ref{app derivation of equations} contains derivation of the difference equations on the modular kernel originating from the consistency relations.
\section{Conformal blocks\label{sec conformal blocks}}
\subsection{General discussion}
Conformal blocks are prime constituents of any conformal field theory (CFT) \cite{BPZ}. In the present paper we only deal with two-dimensional CFT. Symmetries of 2d CFT are encoded into the Virasoro algebra spanned by generators $L_{n}, n\in\mathbb{Z}$ with commutation relations
\begin{eqnarray}
[L_n,L_m]=(n-m)L_{n+m}+\f{c}{12}n(n^2-1)\delta_{n+m,0}\label{Virasoro algebra}
\end{eqnarray}
In fact, the full symmetry algebra of the theory contains two copies of the Virasoro algebra (holomorphic + antiholomorphic). In what follows we will pretend that all operators in the theory are holomorphic. This will have no effect on our conclusions about conformal blocks which are by definition holomorphic objects. However, disregarding antiholomorphic part will greatly lighten the notation.
  
Space of states in CFT is isomorphic to the space of local operators. It is decomposed into the direct sum of the irreducible highest weight representations called Verma modules or conformal families. Highest weight vectors $\ket{\Delta}$ are eigenfunctions of $L_0$ and they are annihilated by all $L_{n}$ with positive $n$
\begin{align}
&L_0\ket{\Delta}=\Delta\nn\\ &L_{n}\ket{\Delta}=0,\quad n>0
\end{align}
Verma module is conveniently organized into levels labeled by a non-negative integer $k$. Vectors of the form 
\begin{align}
&L_{-Y}\ket{\Delta}\equiv L_{-1}^{k_1}L_{-2}^{k_2}\dots\ket{\Delta}
\end{align} 
constitute a basis at level $k$.
Here $Y=\{k_1,k_2,\dots\}$ is a partition of $k$, i.e. $|Y|=k_1+2k_2+\dots = k$. Thus, the operator product expansion (OPE) in 2d CFT can be written in the following form
\begin{eqnarray}
\phi_{\Delta_1}(x)\phi_{\Delta_2}(0)=\sum_{\Delta,Y}C^{\Delta,Y}_{\Delta_1\Delta_2}(x)L_{-Y}\phi_{\Delta}(0)\label{OPE}
\end{eqnarray}
Conformal invariance fixes functions $C^{\Delta,Y}_{\Delta_1\Delta_2}(x)$ up to constants $C^{\Delta}_{\Delta_1\Delta_2}$ independent of partition $Y$
\begin{eqnarray}
C^{\Delta,Y}_{\Delta_1\Delta_2}(x)=C^{\Delta}_{\Delta_1\Delta_2}\be^{\Delta,Y}_{\Delta_1\Delta_2}x^{\Delta-\Delta_1-\Delta_2} 
\end{eqnarray}
One normalizes by definition $\be^{\Delta,\varnothing}_{\Delta_1\Delta_2}=1$, then $C^{\Delta}_{\Delta_1\Delta_2}$ is a  three-point correlation function
\begin{eqnarray}
C^{\Delta_3}_{\Delta_1,\Delta_2}=\left\langle\phi_{\Delta_1}(0)\phi_{\Delta_2}(1)\phi_{\Delta_3}(\infty)\right\rangle\label{structure constants}
\end{eqnarray}
Here a field inserted at infinity is understood as limit $\phi_{\Delta}(\infty)=\lim_{z\to\infty}z^{2\Delta}\phi_{\Delta}(z)$. One should stress that coefficients $\be^{\Delta,Y}_{\Delta_1\Delta_2}$ are completely fixed by the conformal symmetry (requirement that both sides in \eqref{OPE} transform identically). 

 By means of the OPE one can decompose any correlation function into a combination of conformal blocks. Consider four-point correlation function on a sphere \footnote{We want to stress that correlation functions are in fact  \textit{bilinear} combinations of holomorphic \textit{and} antiholomorphic conformal blocks. Thus formula \eqref{full sphere correlator} does not hold literally.}
\begin{eqnarray}
\left\langle\underbrace{\phi_{\Delta_1}(0)\phi_{\Delta_2}(x)}_{\text{OPE}}\underbrace{\phi_{\Delta_3}(1)\phi_{\Delta_4}(\infty)}_{OPE} \right\rangle=\sum_{\Delta} C^{\Delta}_{\Delta_1\Delta_2}C^{\Delta}_{\Delta_3\Delta_4} B_\Delta\begin{bmatrix}
\Delta_2 & \Delta_3\\\Delta_1&\Delta_4
\end{bmatrix}(x)\label{full sphere correlator}
\end{eqnarray}
Here function $B_{\Delta}[\Delta_i](x)$ is the spheric conformal block. It is conveniently depicted graphically as
\begin{figure}[H]
\centering
\begin{tikzpicture}

 \node at (-1,1) {$B_{\Delta}\begin{bmatrix}
 {\Delta}_2&{\Delta}_3\\{\Delta}_1 &{\Delta}_4
 \end{bmatrix}(x)\quad=$};
 \draw[line width=2pt] (1,0.3) node [below] {${\Delta}_1$} -- ++(0.5,0)  -- ++(0.5,0) -- ++(0,1.5) node[left] {${\Delta}_2$} -- ++(0,-1.5) -- ++(0.5,0) node [below] {${\Delta}$} -- ++(0.5,0) -- ++(0,1.5) node[right]{${\Delta}_3$} -- ++(0,-1.5) -- ++ (1,0) node [below] {${\Delta}_4$};
\draw +(6,0.75); 
\end{tikzpicture}
\end{figure}\label{CBs} 
\noindent
 By definition, conformal block is a series in powers of $x$
\begin{eqnarray}
B_{\Delta}\begin{bmatrix}
\Delta_2 & \Delta_3\\\Delta_1&\Delta_4
\end{bmatrix}(x)=x^{\Delta-\Delta_1-\Delta_2}\sum_{n=0}^{\infty}x^nB^{(n)}_{\Delta}\begin{bmatrix}
\Delta_2 & \Delta_3\\\Delta_1&\Delta_4
\end{bmatrix}\label{CB x-expansion}
\end{eqnarray}
where
\begin{eqnarray}
B^{(n)}_{\Delta}\begin{bmatrix}
\Delta_2 & \Delta_3\\\Delta_1&\Delta_4
\end{bmatrix}=\sum_{\substack{Y,Y'\\|Y|=|Y'|=n}}\be^{\Delta,Y}_{\Delta_1\Delta_2}\be^{\Delta,Y}_{\Delta_1\Delta_2}\left\langle L_{-Y}\phi(0)L_{-Y'}\phi(\infty)\right\rangle
\end{eqnarray}
 We emphasize that conformal block \eqref{CB x-expansion} is unambiguously fixed by conformal symmetry. We present here first few terms of the $x$-expansion
\begin{multline}
B_{\Delta}\begin{bmatrix}
\Delta_2 & \Delta_3\\\Delta_1&\Delta_4
\end{bmatrix}(x)=x^{\Delta-\Delta_1-\Delta_2}\left(1+x\frac{\left(\Delta -\Delta _1+\Delta _2\right) \left(\Delta +\Delta _3-\Delta _4\right) }{2 \Delta }+\right.\\\left.x^2\left(\frac{(c+8 \Delta ) \left(\Delta -\Delta _1+\Delta _2\right) \left(1+\Delta -\Delta _1+\Delta _2\right) \left(\Delta
   +\Delta _3-\Delta _4\right) \left(1+\Delta +\Delta _3-\Delta _4\right)}{4 \Delta  \left(c-10 \Delta +2 c \Delta +16
   \Delta ^2\right)}+\dots \right)\right.+\\\left.O(x^3)\right.\Big)\label{CB explicit x-expansion}
\end{multline}

We now turn to the CFT on a torus. Simplest non-trivial example of the toric blocks is the conformal block for the one-point correlation function
\begin{eqnarray}
B_\Delta(\Delta_{ext}|\tau)=Tr_{\Delta}\,q^{L_0-c/24}\label{toric CB},\qquad q=e^{2\pi i \tau}
\end{eqnarray}
or, graphically
\begin{figure}[H]
\centering
\begin{tikzpicture}
\draw[line width=2pt] (2.5,0)--(4,0);
\draw[line width=2pt] (4.7,0) circle (20 pt);
\node[left] at (2,0) {$B_\Delta(\Delta_{ext}|\tau)= $};
\node[above] at (3.2,0) {$\Delta_{ext}$};
\node[above] at (4.7, 22 pt) {$\Delta$};
\end{tikzpicture}
\end{figure}
\noindent Here $\tau$ is the modular parameter of the torus. The trace is taken over the Verma module of primary dimension $\Delta$; $\Delta_{ext}$ stays for the dimension of the external field. Similarly to the spheric case toric conformal block naturally admits expansion in powers of $q$
\begin{multline}
B_{\Delta}(\Delta_{ext}|\tau)=q^{\Delta-c/24}\sum_{n=0}^{\infty}q^{n}B^{(n)}_{\Delta}(\Delta_{ext})=\\q^{\Delta-c/24}\left(1+q\left(\f{\Delta_{ext}(\Delta_{ext}-1)}{2\Delta}+1\right)+q^2\left(\f{(8\Delta+c)\Delta^4_{ext}}{4\Delta(c+2c\Delta-10\Delta+16\Delta^2)}+\dots\right)+O(q^3)\right)\label{toric CB q-expansion}
\end{multline}
Closed expressions for coefficients at arbitrary level $n$ can be obtained via the recently established connection to gauge theories \cite{AGT}(a complete list of references is impossible to give here; for the less known case of the toric conformal block see for example \cite{Alba:2009ya}).  
 Apart from certain special cases to be discussed in the next subsection conformal blocks are best understood as such $x$- or $q$-series expansions.
\subsection{Exactly solvable examples}
However, several exceptions corresponding to certain special choices of parameters exist. First is the case when the central charge equals one and all the external dimensions are equal to $\f1{16}$. Then, the Zamolodchikov's recursive formula \cite{Zamolodchikov:Rec_c,Zamolodchikov:Rec_a} can be solved exactly to give \footnote{In fact, one can ascribe a \textit{different} expression to the conformal block at these values. See \cite{Itoyama:2014dya}.}
\begin{eqnarray}
B_{\Delta}\begin{bmatrix}
1/16 & 1/16\\1/16&1/16
\end{bmatrix}(x)\stackrel{c=1}{=}\f{e^{i\pi \tau\Delta}}{\left(x(1-x)\right)^{1/8}\theta_3(\tau)},\quad \tau = i\f{K(1-x)}{K(x)}\label{Ashkin-Teller}
\end{eqnarray}
where $K(x)$ is the complete elliptic integral of the first kind; $\theta_3(\tau)$ is the standard elliptic theta-constant.  Another exactly solvable example is found when $c=25$ while the all external dimensions $\Delta_i=\f{15}{16}$. This case is quite similar to \eqref{Ashkin-Teller} and we do expose it here.

Toric conformal block can be computed exactly when $c=1$ and $\Delta_{ext}=0$
\begin{eqnarray}
B_{\Delta}(0|\tau)\stackrel{c=1}{=}\f{e^{2\pi i \tau \Delta}}{\eta(\tau)}\label{toric CB exact}
\end{eqnarray}
where $\eta(\tau)$ is the Dedekind eta function. Both our examples of the exact conformal blocks rest on the same simplification occurring in the Zamolodchikov's recurrence formula. In fact, they are essentially identical due to the correspondence between spheric and toric conformal blocks \cite{Poghossian, Hadasz:2009db, Hadasz:2009sw}.

Another important class of exact solutions is found when one of the external dimensions takes specific values corresponding to the degenerate representations of the Virasoro algebra. To describe them efficiently we introduce the Liouville-type parametrization for the central charge and conformal dimensions
\begin{eqnarray}
c=1+6Q^2, \quad Q=b+b^{-1},\quad \Delta(a)=a(Q-a)\label{Liouville notation}
\end{eqnarray} 
Then, degenerate dimensions correspond to the following Liouville momenta
\begin{eqnarray}
\Delta_{\text{deg}}=a_{\text{deg}}(Q-a_{\text{deg}}),\qquad a_{\text{deg}}=-\f{nb}{2}-\f{mb^{-1}}{2},\quad n,m\ge 0 \label{degenerate dimensions}
\end{eqnarray}
When conformal dimension of a field is degenerate its Verma module contains a singular vector. As a consequence, correlation functions and conformal blocks involving a degenerate field satisfy certain differential equations.

We illustrate this at the simplest and most important for our purposes case of the degenerate field with the Liouville momenta $-b/2$. From commutation relations of the Virasoro algebra \eqref{Virasoro algebra} it is easy to check that vector
\begin{eqnarray}
\ket{\text{sing}}=\left(L_{-1}^2+b^2L_{-2}\right)\ket{\Delta(-b/2)}
\end{eqnarray}
is indeed a singular vector, i.e. it is annihilated by all $L_{n}$ with $n>0$ (this is only non-trivial for $n=1,2$). In effect, any correlation function involving $\ket{\text{sing}}$ must vanish and for four-point correlation function
\begin{eqnarray}
C\begin{bmatrix}
\Delta(-b/2) & \Delta_3\\ \Delta_1 & \Delta_4
\end{bmatrix}(x)=\left\langle\phi_{\Delta_1}(0)\phi_{\Delta(-b/2)}(x)\phi_{\Delta_3}(1)\phi_{\Delta_4}(\infty)\right\rangle\label{degenerate correlator}
\end{eqnarray}
presence of a singular state produces equation
\begin{eqnarray}
\left(b^{-2} x(1-x)\p_x^2+(2x-1)\p_x+\Delta(-b/2)+\f{\Delta_1}{x}-\f{\Delta_3}{x-1}-\Delta_4\right)C\begin{bmatrix}
\Delta(-b/2) & \Delta_3\\ \Delta_1 & \Delta_4
\end{bmatrix}(x)=0
\end{eqnarray}
Two independent solutions to this equation can be chosen as
\begin{align}
&B_{a_1-b/2}\begin{bmatrix}
-b/2 & a_3\\a_1&a_4
\end{bmatrix}(x)=x^{ba_1}(1-x)^{ba_3}{}_2F_1(A,B;C|x)\nn\\ {}
&B_{a_1+b/2}\begin{bmatrix}
-b/2 & a_3\\a_1&a_4
\end{bmatrix}(x)=x^{b(Q-a_1)}(1-x)^{ba_3}{}_2F_1(1+A-C,1+B-C;2-C|x)\label{degenerate CBs}
\end{align}
where
\begin{eqnarray}
A=-\f{b^2}{2}+b(a_1+a_3-a_4),\quad B = b(a_1+a_3+a_4)+2\Delta(-b/2),\quad C=-b^2+2ba_1\label{A,B,C}
\end{eqnarray}
One can verify in a first few orders of $x$-expansion that expressions in the r.h.s. of \eqref{degenerate CBs} coincide with the general formula \eqref{CB explicit x-expansion} upon substituting to the latter parametrization \eqref{Liouville notation} and specifying $a=a_1\pm b/2, a_2=-b/2$.

Thus, for correlation function \eqref{degenerate correlator} only two conformal blocks \eqref{degenerate CBs} in decomposition \eqref{full sphere correlator} are relevant. This is a manifestation of the fusion rules for the OPE involving degenerate fields. Particularly
\begin{eqnarray}
\phi_{a}\times\phi_{-b/2}=\phi_{a+b/2}+\phi_{a-b/2}\label{fusion rules}
\end{eqnarray}
In words, only operators of momenta $a\pm b/2$ may have non-vanishing coefficients in the OPE of fields $\phi_{a},\phi_{-b/2}$. Hence the space of conformal blocks is two-dimensional and functions \eqref{degenerate CBs} can be chosen as a basis.

One can see that our list of explicitly known conformal blocks is quite skimpy and only covers very special circumstances. Different approaches towards understanding the generic Virasoro conformal block may be chosen. One is provided by the AGT correspondence \cite{AGT, Wyllard:2009hg, Mironov:2009qt, Mironov:2009by,Alba:2010qc, Belavin:2011js, Marshakov:2009gs, Morozov:2013rma, Mironov:2013oaa} which in particular relates conformal blocks to the partition functions of supersymmetric gauge theories, where a lot of explicit computations can be carried using localization technique \cite{Pestun:2007rz, Nekrasov:2003rj, Nekrasov:2002qd}.  Another is given by the matrix model representation of conformal blocks \cite{Dotsenko:1984nm, Mironov:2010zs, Mironov:2009ib, Mironov:2010xs, Mironov:2010ym}. Intriguing is the recently established connection between conformal blocks and the theory of Painlev\'e equations \cite{Gamayun:2012ma, Litvinov:2013sxa}. Advances are also being made purely withing the CFT framework \cite{Perlmutter:2015iya}. In the present paper we take yet another possible direction. We wish to study one particular aspect of conformal blocks. Namely, their behavior under the fusion and modular transformations to be described in the next section. 
\section{Fusion and modular transformations\label{sec modular transformations}}
\subsection{General discussion}
In order to obtain decomposition \eqref{full sphere correlator} we have chosen a particular pairing of the fields in the correlation function. Namely, we fused fields with dimensions $\Delta_1,\Delta_2$ and $\Delta_3,\Delta_4$.
Fusing fields in a different way would result in a different basis for conformal blocks. For example
\begin{eqnarray}
\left\langle\underbrace{\phi_{\Delta_1}(0)\phi_{\Delta_4}(\infty)}_{\text{OPE}}\underbrace{\phi_{\Delta_2}(x)\phi_{\Delta_3}(1)}_{OPE} \right\rangle=\sum_{\Delta} C^{\Delta}_{\Delta_1\Delta_4}C^{\Delta}_{\Delta_2\Delta_3} B^t_\Delta\begin{bmatrix}
\Delta_2 & \Delta_3\\\Delta_1&\Delta_4
\end{bmatrix}(x)\label{full sphere correlator t}
\end{eqnarray}
Function $B^t_{\Delta}[\Delta_i](x)$ appearing in this decomposition is called t-channel conformal block (in contranst to \eqref{CBs} which is called s-channel conformal block) and depicted as
\begin{figure}[H]
\centering
\begin{tikzpicture}
 \node at (-1,0.7) {$B^{t}_{\Delta}\begin{bmatrix}
 {\Delta}_2&{\Delta}_3\\{\Delta}_1&{\Delta}_4
 \end{bmatrix}(x)\quad=$};
\draw[line width=2pt] (1,0)node [below] {${\Delta}_1$} -- ++(1.5,0)--++(1.5,0) node[below]{${\Delta}_4$}--++(-1.5,0)-- ++(0,0.75) node[below right]{${\Delta}$} -- ++(-0.75,0.75)node[left]{${\Delta}_2$} -- ++(0.75,-0.75)--++(0.75,0.75) node[right]{${\Delta}_3$};
\end{tikzpicture}
\end{figure}
Throughout this section we introduce additional labels $s$ and $t$ to differentiate between s- and t-channel blocks. In the subsequent sections we only use s-channel blocks and hence, drop the superscript. There is a simple relation between s- and t- channel conformal blocks
\begin{eqnarray}
B^{t}_{\Delta}\begin{bmatrix}
 {\Delta}_2&{\Delta}_3\\{\Delta}_1&{\Delta}_4
 \end{bmatrix}(x)=B^{s}_{\Delta}\begin{bmatrix}
  {\Delta}_2&{\Delta}_1\\{\Delta}_3&{\Delta}_4
  \end{bmatrix}(1-x)\label{s to t channel}
\end{eqnarray}
From asymptotic near $x=0$ \eqref{CB x-expansion} one sees that s-channel conformal blocks are linearly independent. Then, from \eqref{s to t channel} we conclude that the t-channel conformal blocks are also linearly independent. Therefore, decompositions into s- and t-channels \eqref{full sphere correlator}, \eqref{full sphere correlator t} are simply decompositions in different bases. Hence these bases are related by some transformation matrix
\begin{eqnarray}
B^{s}_{\Delta}\begin{bmatrix}
 {\Delta}_2&{\Delta}_3\\{\Delta}_1&{\Delta}_4
 \end{bmatrix}(x)=\sum_{\Delta'}F_{\Delta\Delta'}\begin{bmatrix}
 \Delta_2&\Delta_3\\\Delta_1&\Delta_4
 \end{bmatrix}B^{t}_{\Delta}\begin{bmatrix}
 {\Delta}_2&{\Delta}_3\\{\Delta}_1&{\Delta}_4
 \end{bmatrix}(x)\label{FK definition}
\end{eqnarray}
Here the summation is performed over the spectrum of primary fields' dimensions. One might question formula \eqref{FK definition} when the spectrum is continuous, but its validity is a common belief confirmed from various perspectives. Transformation from t- to s-channel is sometimes called fusion. Hence, we call matrix $F_{\Delta\Delta'}[\Delta_i]$ \textit{the fusion kernel}.

From relation between s- and t-channel \eqref{s to t channel} we see that the fusion kernel not only describes fusion transformation, but rather monodromy properties of the s-channel conformal block itself
\begin{eqnarray}
B^{s}_{\Delta}\begin{bmatrix}
 {\Delta}_2&{\Delta}_3\\{\Delta}_1&{\Delta}_4
 \end{bmatrix}(x)=\sum_{\Delta'}F_{\Delta\Delta'}\begin{bmatrix}
 \Delta_2&\Delta_3\\\Delta_1&\Delta_4
 \end{bmatrix}B^{s}_{\Delta}\begin{bmatrix}
 {\Delta}_2&{\Delta}_1\\{\Delta}_3&{\Delta}_4
 \end{bmatrix}(1-x)\label{FK as monodromy}
\end{eqnarray}
and therefore contains non-trivial information about the structure of the conformal blocks.

Correlation functions on a torus must be invariant under the action of $SL(2,\mathbb{Z})$ generated by $T:\tau\to\tau+1$ and $S:\tau\to-1/\tau$. As a consequence, toric conformal blocks form a representation of the modular group. $T$-transformation acts diagonally and can be read off directly from definition \eqref{toric CB}
\begin{eqnarray}
 B_{\Delta}(\Delta_{ext}|\tau+1)=q^{\Delta-c/24}B_{\Delta}(\tau)
\end{eqnarray} 
The action of $S$ is non-trivial and close in spirit to the fusion transformation \eqref{FK definition}
\begin{eqnarray}
B_{\Delta}(\Delta_{ext}|-1/\tau)=\sum_{\Delta'}M_{\Delta\Delta'}(\Delta_{ext})\,B_{\Delta'}(\Delta_{ext}|\tau)
\end{eqnarray}
We call matrix $M_{\Delta\Delta'}$ \textit{the toric modular kernel} or simply the modular kernel.

The present paper is mainly concerned with the toric modular kernel. It is simpler then the fusion kernel (already by counting the number of parameters) and allows for a clearer exposition while catching all the important features. Though might be tedious, generalization to the spheric case should be straightforward. In \cite{PT1, PT2, PT3} the authors proposed formulas  describing fusion and modular kernels for generic conformal blocks . Before presenting and discussing these formulas here we first illustrate fusion and modular transformations on the exactly solvable examples.

In the following for the sake of brevity we often use the term modular transformations for both, fusion transformations of the spheric and S-transformations of the toric blocks. Hopefully, this is not to be a source of confusion.
\subsection{Exactly solvable examples}
In practice it turns out to be more convenient to use the Liouville-type parametrization \eqref{Liouville notation}. Hence we rewrite \eqref{FK as monodromy} as
\begin{eqnarray}
B_{a}\begin{bmatrix}
 {a}_2&a_3\\a_1&a_4
 \end{bmatrix}(x)=\int da' F_{aa'}\begin{bmatrix}
 a_2&a_3\\a_1&a_4
 \end{bmatrix}B_{a'}\begin{bmatrix}
 a_2&a_1\\a_3&a_4
 \end{bmatrix}(1-x)\label{FK as monodromy Liouville}
\end{eqnarray} 
Note also that we have dropped s-channel superscript. Here and in the sequel spheric conformal block without superscript always refers to s-channel conformal block.	

Consider first Zamolodchikov's example \eqref{Ashkin-Teller} which is a Gaussian function of the Liouville momenta
\begin{eqnarray}
B_{a}\begin{bmatrix}
i/4&i/4\\i/4&i/4
\end{bmatrix}(x)\stackrel{c=1}{=}\f{e^{-i\pi\tau a^2}}{\left(x(1-x)\right)^{1/8}\theta_3(
\tau)}
\end{eqnarray}
Note that transformation $x\to1-x$ renders $\tau\to-1/\tau$. Then, using $\theta_3(-1/\tau)=\s{-i\tau}\theta_3(\tau)$ and performing simple integration one shows that
\begin{eqnarray}
\f{e^{-i\pi\tau a^2}}{\left(x(1-x)\right)^{1/8}\theta_3(
\tau)}=\int da' e^{2\pi i aa'}\f{e^{i\pi a'^2/\tau}}{\left(x(1-x)\right)^{1/8}\theta_3(-1/
\tau)}
\end{eqnarray}
Therefore, in the case of the unit central charge and all the external momenta equal to $i/4$ (corresponding to the dimensions equal to $1/16$) the fusion transformation is simply the Fourier transformation
\begin{eqnarray}
F_{aa'}\begin{bmatrix}
i/4&i/4\\i/4&i/4
\end{bmatrix}\stackrel{c=1}{=}e^{2\pi i aa'}
\end{eqnarray}
In the other case when the Zamolodchikov's recursion relation for the spheric block can be solved the result is no different, fusion transformation is the pure Fourier transformation. 

As already mentioned, formulas \eqref{Ashkin-Teller} and \eqref{toric CB exact} are basically equivalent and so are the transformation properties of these conformal blocks. Namely, using $\eta(-1/\tau)=\s{-i\tau}\eta(\tau)$ one shows that
\begin{eqnarray}
\f{e^{-2i\pi\tau a^2}}{\eta(
\tau)}=\int da' e^{4\pi i aa'}\f{e^{2i\pi a'^2/\tau}}{\eta(-1/\tau)}
\end{eqnarray}
Hence,
\begin{eqnarray}
M_{aa'}(0)\stackrel{c=1}{=}e^{4\pi i aa'}
\end{eqnarray}
which is again the pure Fourier transform with a slightly different normalization convention.

Now let us consider conformal blocks with a degenerate field of momentum $-b/2$ \eqref{degenerate CBs}. Recall that due to the fusion rules \eqref{fusion rules} the space of conformal blocks is two-dimensional. Basis s-channel conformal blocks are given by formulas \eqref{degenerate CBs}. Let us introduce notation
\begin{eqnarray}
B^s_\pm=B^s_{a_1\pm b/2}\begin{bmatrix}
-b/2&a_3\\a_1&a_4
\end{bmatrix}(x) 
\end{eqnarray} 
Note that permutation $a_1\leftrightarrow a_3$ acts on parameters $A,B,C$ \eqref{A,B,C} in the following way
\begin{eqnarray}
A\to A,\quad B\to B,\quad C\to A+B-C+1
\end{eqnarray}
Therefore, in this case the fusion kernel is nothing else than the matrix of connection coefficients for the hypergeometric function  
\begin{eqnarray}
\begin{pmatrix}
B^s_+\\B^s_-
\end{pmatrix}=F\begin{pmatrix}
B^t_+\\B^t_-
\end{pmatrix},\qquad\qquad F\begin{bmatrix}
-b/2&a_3\\a_1&a_4
\end{bmatrix}=\left(
\begin{array}{cc}
 \frac{\Gamma (2-C) \Gamma (A+B-C)}{\Gamma (1+A-C) \Gamma (1+B-C)}
   & \frac{\Gamma (2-C) \Gamma (-A-B+C)}{\Gamma (1-A) \Gamma (1-B)}
   \\
 \frac{\Gamma (A+B-C) \Gamma (C)}{\Gamma (A) \Gamma (B)} &
   \frac{\Gamma (C) \Gamma (-A-B+C)}{\Gamma (-A+C) \Gamma (-B+C)}
   \\
\end{array}
\right) \label{degenerate fusion matrices}
\end{eqnarray}
\section{Toric modular kernel from representation theory of $\mathcal{U}_q(sl_2)$\label{sec PT formula}}
In paper \cite{PT3} formula for the modular kernel of the generic toric conformal block was presented in the following form \footnote{We warn the reader of notational difference with the original paper. The exact relation is $\al=ip_a,\al'=ip_b,\mu=Q/2+ip_e,c_b=iQ/2, s_b(z)=S_b(Q/2+iz)$.}
\begin{eqnarray}
\boxed{\mathcal{M}_{\al\al'}(\mu)=\f{2^{3/2}}{i}\f{\sin{2\pi b\al'}\sin{2\pi b^{-1}\al'}}{S_b(\mu)}\int_{\mathcal{C}} d\xi \f{S_b(\al'+\f\mu2+\xi)S_b(\al'+\f\mu2-\xi)}{S_b(\al'+Q-\f\mu2+\xi)S_b(\al'+Q-\f\mu2-\xi)}e^{-4\pi i \al \xi}}\label{MK by PT}
\end{eqnarray}
Here $S_b(z)$ is a special function we call the double sine function. It is described in appendix \ref{app special functions}. Several remarks are in order. \\
1) Parameters $\al,\al'$ in \eqref{MK by PT} are shifted Liouville momenta for the internal dimensions
\begin{eqnarray}
\al=a-Q/2,\quad \al'=a'-Q/2
\end{eqnarray}
while $\mu$ is the original Liouville momenta for the external dimension. Thus
\begin{eqnarray}
\Delta=\f{Q^2}{4}-\al^2,\quad \Delta'=\f{Q^2}{4}-\al'^2,\quad \Delta_{ext}=\mu(Q-\mu) 
\label{al parameters}
\end{eqnarray}
2) Modular kernel \eqref{MK by PT} is presented in a special basis of conformal blocks, with normalization different from standard \eqref{toric CB}, \eqref{toric CB q-expansion}. The exact relation is as follows. Define renormalization factor for the chiral vertex
\begin{eqnarray}
V(a_1,a_2;a_3)=\frac{\Gamma_b(2a_1)\Gamma_b(2a_2)\Gamma_b(2Q-2a_3)}{\Gamma_b(2Q-a_1-a_2-a_3)\Gamma_b(a_1+a_2-a_3)\Gamma_b(a_1-a_2+a_3)\Gamma_b(-a_1+a_2+a_3)}\label{normalization}
\end{eqnarray}
Here $\Gamma_b(z)$ is a special function we call the double gamma function \ref{app special functions}. Toric conformal block contains a single vertex and thus is renormalized as
\begin{eqnarray}
\mathcal{B}_{a}(\mu|\tau)=V(\mu,a;a)B_{a}(\mu|\tau)
\end{eqnarray}
Hence, modular kernel $\mathcal{M}_{\al\al'}(\mu)$ in \eqref{MK by PT} is related to the modular kernel $M_{\al\al'}(\mu)$ in the standard normalization of conformal blocks \eqref{toric CB} as
\begin{eqnarray}
\mathcal{M}_{\al\al'}(\mu)=\f{V(\mu,\al+Q/2;\al+Q/2)}{V(\mu,\al'+Q/2;\al'+Q/2)} M_{\al\al'}(\mu) \label{toric normalization}
\end{eqnarray}
3) For generic $c>1$ integrand in \eqref{MK by PT} contains four infinite half-lines of poles and integration contour $\mathcal{C}$ is chosen to  maneuver between them in a specific way. For certain values of parameters some of these poles  can overlap and  merge and formula \eqref{MK by PT} needs  a completion. In some cases this issue is not resolvable. For example, for $c\le1$ (and thus for all minimal models) the formula is not valid.

\section{Main result: series expansion of the toric modular kernel\label{sec main result}}
Main result of the present work is a series representation of formula \eqref{MK by PT}. With the same notation and normalization conventions the series expansion reads

\begin{empheq}[box=\fbox]{multline}
\mathcal{M}_{\al\al'}(\mu)=2^{3/2}e^{4\pi i \al\al'}e^{2\pi i (\mu\al+(Q-\mu)\al')}e^{i\pi\mu(Q-\mu)/2}\sin{2\pi b\al'}\sin{2\pi b^{-1}\al'}\times \\\sum_{n,m=0}^{\infty}e^{4\pi i nb\al}e^{4\pi i mb\al'}e^{2\pi i nmb^2}\prod_{k=1}^n \f{e^{2\pi i k b^2}-e^{2\pi i b(b-\mu)}}{e^{2\pi i k b^2}-1}\prod_{l=1}^m \f{e^{2\pi i l b^2}-e^{2\pi i b\mu}}{e^{2\pi i l b^2}-1}\times (b\to b^{-1})\\+(\al'\to-\al')\label{mk}
\end{empheq}

Or, less schematically, but more bulky
\begin{align}
\mathcal{M}_{\al\al'}(\mu)=2^{5/2}e^{2\pi i \mu\al}e^{i\pi\mu(Q-\mu)/2}&\sin{2\pi b\al'}\sin{2\pi b^{-1}\al'}\times \nn\\\sum_{n,m,\tilde{n},\tilde{m}=0}^{\infty}\Bigg(e^{4\pi i (nb+\tilde{n}b^{-1})\al} &\cos{\left(2\pi\al'(2\al+Q-\mu+2mb+2\tilde{m}b^{-1})\right)}\times \nn\\ {}e^{2\pi i nmb^2}&\prod_{k=1}^n \f{e^{2\pi i k b^2}-e^{2\pi i b(Q-\mu)}}{e^{2\pi i k b^2}-1}\prod_{l=1}^m \f{e^{2\pi i l b^2}-e^{2\pi i b\mu}}{e^{2\pi i l b^2}-1}\times\nn\\ {} & e^{2\pi i \tilde{n}\tilde{m}b^{-2}}\prod_{\tilde{k}=1}^{\tilde{n}} \f{e^{2\pi i \tilde{k} b^{-2}}-e^{2\pi i b^{-1}(Q-\mu)}}{e^{2\pi i \tilde{k} b^{-2}}-1}\prod_{\tilde{l}=1}^{\tilde{m}} \f{e^{2\pi i \tilde{l} b^{-2}}-e^{2\pi i b^{-1}\mu}}{e^{2\pi i \tilde{l} b^{-2}}-1} \Bigg)
\end{align}
Several comments are in order.\\
1) This representation is valid in any domain of the parameter space where the series is convergent. For example, the default setup in the Liouville theory is $c\ge1$ restricting $Q\in\mathbb{R}$ while ${\Delta\ge(c-1)/24\in\mathbb{R}}$ imposing $\al\in i\mathbb{R},\mu\in \f{Q}{2}+i\mathbb{R}$. Then, the series expansion is convergent for $b\in \mathbb{R},\al\in i\mathbb{R}_+$. The formula can be trivially continued to $\al\in -i\mathbb{R}_+$ by requiring that $\cm_{\al\al'}(\mu)$ is an even function of $\al$. In fact, the domain of validity of the series representation seems to be the same as for the integral formula. For instance, for the minimal models corresponding to $c<1$ we have to pick imaginary $b$ and then exponents $e^{4\pi i nb\al}$ refuse to be small rendering the series not convergent.\\
2) The necessity to add a second term with $\al'\to-\al'$ in \eqref{mk} is somewhat formal. For instance, since the conformal block can only depend on $\al$ through a conformal dimension $\Delta=Q^2/4-\al^2$ it must be an even function of momentum $\al$. Therefore, for the transformations of the conformal blocks $\al'$-odd part of the modular kernel is not essential.
 
  In the sequel we derive this series expansion both, directly from formula \eqref{MK by PT} and by solving consistency equations which are satisfied by the modular kernel. The latter way, though not truly independent, may be viewed as an additional verification of the integral formula. It is also fruitful for some border cases not described by \eqref{MK by PT}, such as $c=1$.

\section{Series representation as non-perturbative completion\label{sec non-perturbative discussion}}
Formulas for the \textit{generic} modular or fusion kernel discussed in the present work are quite formal. As already mentioned, they are not derived from the modular properties of the \textit{generic} conformal blocks. Rather, only properties of the very limited family of the \textit{degenerate} conformal blocks together with some indirect arguments (consistency conditions) were used to produce formulas \eqref{MK by PT},\eqref{mk}. In other words, we know the transformation laws but we can not appropriately describe the objects subjected to the transformations. 

Attempts to clear this question were made in papers \cite{GMMpert, Lerda, Billo:2013fi, Nemkov1}. Technically, the main problem is that by definition the spheric conformal block is a series expansions in powers of $x$, while fusion transformation changes $x\to1-x$. Likewise, the toric conformal block is a series in $e^{2\pi i \tau}$ while the modular transformation changes $\tau\to-1/\tau$. In these expansions properties under the modular transformations are concealed. Certain techniques were exploited in works \cite{GMMpert, Lerda, Billo:2013fi, Nemkov1} to rewrite the conformal block as the series in powers of $\al^{-1}$ with coefficients being exact functions of $x$ or $\tau$. Then, the modular properties of each coefficient become manifest and one can construct the modular kernel perturbatively in $\al^{-1}$. The expected result was
\begin{eqnarray}
M_{\al\al'}(\mu)=e^{4\pi i \al\al'+O(\al^{-1},\al'^{-1})}
\end{eqnarray}
However, it appeared that the perturbative (in the above sense of the large $\al$ expansion) conformal blocks transform exactly according to the Fourier transformation
\begin{eqnarray}
M_{\al\al'}(\mu)=e^{4\pi i \al\al'}
\end{eqnarray}
with all the perturbative corrections $O(\al^{-1},\al'^{-1})$ vanishing.

 We see \footnote{Apart from a slight normalization difference  due to renormalization given below formula \eqref{MK by PT}} that the Fourier kernel is the leading asymptotic in formula \eqref{mk} at large $\al,\al'$, while terms including $e^{4\pi i b\al}, e^{4\pi i b\al'}$ are exponentially small. The claim of \cite{GMMpert, Lerda, Billo:2013fi, Nemkov1} was that perturbatively the modular transform is exactly the Fourier transform. From this point of view formula \eqref{mk} describes non-perturbative corrections. It is therefore important and unresolved question to directly relate these corrections to the conformal blocks. A possibility is, that the very definition of conformal blocks requires non-perturbative completion \cite{Itoyama:2014dya}.

\section{Modular kernel asymptotic expansion from consistency requirements\label{sec consistency constraints}}
\subsection{Pentagon identity and difference equations on the fusion kernel \label{pentagon section}}
As is well known the fusion kernel satisfies the pentagon identity
\begin{eqnarray}F_{p_1 q_2}\begin{bmatrix}
a_2&a_3\\a_1&p_2
\end{bmatrix}F_{p_2 q_1}\begin{bmatrix}
q_2&a_4\\a_1&a_5
\end{bmatrix}=
\sum_{l} F_{p_2 l}\begin{bmatrix}
a_3&a_4\\p_1&a_5
\end{bmatrix}F_{p_1 q_1}\begin{bmatrix}
a_2&l\\a_1&a_5
\end{bmatrix}F_{l q_2}\begin{bmatrix}
a_2&a_3\\q_1&a_4
\end{bmatrix}\label{pentagon}
\end{eqnarray}
This formula follows from the requirement of consistency for the fusion transformations of the 5-point conformal blocks. Labels $a_{1-5}$ indicate the external momenta in these 5-point blocks. When one of these momenta, say $a_2$ is set to a degenerate value $a_2=-b/2$ two interesting phenomena happen:\\[2pt]
1) Three out of five fusion matrices entering the pentagon identity become degenerate and therefore known explicitly.  The two other matrices are left with generic values of parameters. This converts polynomial equation \eqref{pentagon} into a linear equation on the \textit{generic} fusion matrices with \textit{degenerate} fusion matrices playing the role of coefficients.\\
2) Due to fusion rules \eqref{fusion rules} generally continuous range of summation over momenta $l$ is restricted to just two values $l=q_1\pm b/2$.  

This turns the pentagon identity into a second order linear homogeneous difference equation. One can attempt to solve this equation directly. In the previous work \cite{Nemkov2} we took this route and provided a way to recursively find coefficients in the expansion of the type \eqref{mk} for the fusion kernel. In the current work we continue to develop this approach at the example of the toric modular kernel. This case appears to be simpler and allows for more complete understanding.
\subsection{Consistency constraints on the modular kernel}
Consistency conditions in the fusion algebra not only restrict the genus zero transformations but also intertwine them with the genus one transformations \cite{MS, MSlecturesRCFT}. This aspect is perhaps less known and we give a derivation in appendix \ref{app derivation of equations}. As a result one obtains the following set of difference equations on the 	modular kernel
\begin{empheq}[box=\fbox]{gather}
\left(\f{\sin{\pi b(2\al+\mu)}}{\sin{2 \pi b\al}}e^{\f{b}{2}\p_{\al}}+\f{\sin{\pi b(2\al-\mu)}}{\sin{2 \pi b\al}}e^{-\f{b}{2}\p_{\al}}\right)\cm_{\al\al'}(\mu)=2\cos{2\pi b\al'}\,\cm_{\al\al'}(\mu)\label{maineq a}\\
\left(e^{-\f{b}{2}\p_{\al'}}\f{\sin{\pi b(2\al'+\mu)}}{\sin{2 \pi b\al'}}+e^{\f{b}{2}\p_{\al'}}\f{\sin{\pi b(2\al'-\mu)}}{\sin{2 \pi b\al'}}\right)\cm_{\al\al'}(\mu)=2\cos{2\pi b\al}\,\cm_{\al\al'}(\mu)\label{maineq a'}\\
\f1{\sin{2\pi b\al}}\left(e^{\f{b}2\p_\al}-e^{-\f{b}2\p_\al}\right)\cm_{\al\al'}(\mu)=2e^{b\p_\mu}\cm_{\al\al'}(\mu)\label{maineq mu}
\end{empheq}
and we have switched to the same notation in which formulas \eqref{MK by PT}, \eqref{mk} are written. In paper \cite{GMMnonpert} it was shown that these equations also follow in the language of matrix models from the formalism of check operators.

Let us discuss these equations. First of them is a second-order difference homogeneous equation on the modular kernel $\cm_{\al\al'}(\mu)$ with shifts in internal momentum $\al$. The second equation is the counterpart with shifts in the other internal momentum, $\al'$. It can be derived from the first and the property that $\cm_{\al\al'}(\mu)$ squares to unity. The third equation involves shifts in external momentum $\mu$. Hereby, we have three equations for the modular kernel which depends on three parameters. 

The first and the second equations are second-order and therefore admit two linearly independent solutions. Furthermore, we deal with the difference equations with $b/2$-valued shifts.  Therefore, solution is only determined up to a function $f(\al,\al',\mu)$ which is $2b^{-1}$-periodic in all parameters. 

These ambiguities are artifacts of our parametrization \eqref{Liouville notation}. Both, $\al$ and $-\al$ correspond to the same conformal dimension $\Delta=\f{Q^2}{4}-\al^2$. Since the modular kernel should be in fact a function of $\Delta$ we shall choose $\cm_{\al\al'}(\mu)$ to be an even function of $\al,\al'$. This eliminates the excess of solutions related to the degree of the equations.  Likewise, the modular kernel should only depend on the central charge $c=1+6(b+b^{-1})^2$ and not on $b$ separately. This requirement fixes undetermined $2b^{-1}$-periodic function $f(\al,\al',\mu)$ up to a function $g(\al,\al',\mu)$ which is both $2b^{-1}$- and $2b$-periodic in all parameters. For generic irrational $b$ this obliges function $g(\al,\al',\mu)$ to be constant. And even this constant can be fixed from relation $\cm^2=1$. Thus, for a generic irrational $b$ the modular kernel is uniquely identified as the solution to equations \eqref{maineq a},\eqref{maineq a'},\eqref{maineq mu} with the desired symmetry properties. 
\subsection{Solutions to consistency constraints}

For generic irrational $b$ our final result is expressed by formula \eqref{mk} which was stated before. In the present section we derive this expression from the difference equations \eqref{maineq a},\eqref{maineq a'},\eqref{maineq mu}. Formula \eqref{mk} is valid for generic irrational values of $b$, but not applicable to say $b\in \mathbb{N}$. We first discuss these special cases. 

\subsubsection{Special values of $b$}
Probably, the two most interesting examples which do not correspond to the irrational values of $b$ is $c=1$ for $b=i$ and $c=25$ for $b=1$. From the perspective of the difference equations these two cases are basically equivalent via map $b\to ib$ (and similarly for momenta $\al,\al',\mu$) and we will discuss here only $b=1$. Neither of formulas \eqref{MK by PT},\eqref{mk} is valid for $b=1$. However, equations \eqref{maineq a},\eqref{maineq a'},\eqref{maineq mu} are valid and moreover greatly simplify.

To see this consider, for example, the first equation. It contains shifts in $\al$ by the value equal to $b/2$. At the same time, coefficients in this equation are $b^{-1}/2$-periodic functions of $\al$. Therefore, for $b=1$ the period of the coefficients coincide with the difference step. This allows to solve the equation as if it had constant coefficients
\begin{eqnarray}
\cm_{\al\al'}(\mu)=\left(m(\al,\al',\mu)\right)^{2\al}\times f(\al,\al',\mu)
\end{eqnarray}
where function $m(\al,\al',\mu)$ is determined from the quadratic equation
\begin{eqnarray}
\f{\sin{\pi(2\al+\mu)}}{\sin{2 \pi \al}}m(\al,\al',\mu)+\f{\sin{\pi (2\al-\mu)}}{\sin{2 \pi \al}}m^{-1}(\al,\al',\mu)=2\cos{2\pi \al'}
\end{eqnarray}
to be
\begin{eqnarray}
m(\al,\al',\mu)=\f{\cos{2\pi\al'}\sin{2\pi\al}\pm \s{\cos^2{2\pi\al'}\sin^2{2\pi\al}-\sin{\pi(2\al+\mu)}\sin{\pi(2\al-\mu)}}}{\sin{\pi(2\al+\mu)}}
\end{eqnarray}
while function $f(\al,\al',\mu)$ is arbitrary function 1-periodic in $\al$. In contrast to the generic case this function can not be fixed within our approach. One needs other arguments to find it. 

Our discussion of the $c=25$ modular kernel could be translated with little changes to the $c=1$ fusion kernel. This question was investigated in detail in paper \cite{Nemkov2}. On the other hand, for $c=1$ fusion kernel exists another formula provided by recently discovered connection with the theory of Painlev\'e equations \cite{Iorgov}. Despite the fact that approach of \cite{Iorgov} also reduced to solving certain difference equations the authors were able to fix the ambiguity by matching new general formula with  particular examples found earlier within the Painlev\'e theory. Explicit expressions for $c=1$ fusion kernel in \cite{Nemkov2} and \cite{Iorgov}  look very different (the latter significantly more complicated), but it was shown that their non-periodic parts coincide. 

To sum up, approach based on difference equations  \eqref{maineq a},\eqref{maineq a'},\eqref{maineq mu} is still fruitful for border cases like $c=1$ which are not described by formulas \eqref{MK by PT},\eqref{mk}. Moreover, solutions are much simpler than in the generic case. However, in order to remove ambiguities one is forced to resort to some other arguments. Addressing this issue goes beyond the scope of the present work. 
\subsubsection{Generic irrational values of $b$}
Equation \eqref{maineq a} can be rewritten in the following form
\begin{multline}
\left(e^{-i\pi b\mu}e^{\f{b}2\p_\al}+e^{i\pi b\mu}e^{-\f{b}2\p_\al}-2\cos{2\pi b\al'}\right)\mathcal{M}_{\al\al'}(\mu)\\=e^{4\pi i b \al }\left(e^{i\pi b\mu}e^{\f{b}2\p_\al}+e^{-i\pi b\mu}e^{-\f{b}2\p_\al}-2\cos{2\pi b\al'}\right)\mathcal{M}_{\al\al'}(\mu)\label{maineq a reformulated}
\end{multline}
which suggests to look for a solution of this equation in terms of the formal (trans-series) expansion in powers of $e^{4\pi i b \al}$
\begin{eqnarray}
\mathcal{M}_{\al\al'}= \sum_{n=0}^{\infty}e^{4\pi in b\al} \tilde{\mathcal{M}}_{\al\al'}^n \label{trans-series ansatz}
\end{eqnarray}
where each coefficient $\tilde{\mathcal{M}}_{\al\al'}^n$ is a perturbative series in $\al^{-1}$ (as we will soon show, each $\tilde{\cm}^n_{\al\al'}$ is basically independent of $\al$). We have seen that in the standard region of the Liouville parameter space exponents $e^{4\pi i b \al}$ are indeed small and suitable for perturbative expansion.

Let us look at the zeroth order equation
\begin{eqnarray}
\left(e^{-i\pi b\mu}e^{\f{b}2\p_\al}+e^{i\pi b\mu}e^{-\f{b}2\p_\al}-2\cos{2\pi b\al'}\right)\tilde{\cm}^0_{\al\al'}(\mu)=0
\end{eqnarray}
It is easy to find two linearly independent solutions 
\begin{eqnarray}
\tilde{\mathcal{M}}^0_{\al\al'}(\mu)=e^{\pm4\pi i \al\al'}e^{2\pi i \al\mu}
\end{eqnarray}
In the end we are interested in the even combination of these solutions but it is simpler to pick one exponent and impose the symmetry requirement afterwards. We thus chose zeroth order term as
\begin{eqnarray}
\tilde{\mathcal{M}}^{0}_{\al\al'}(\mu)=e^{4\pi i \al\al'}e^{2\pi i \al\mu}\label{zeroth order choice}
\end{eqnarray}
It appears convenient to factor this contribution out of the whole series
\begin{eqnarray}
\mathcal{M}_{\al\al'}(\mu)=e^{4\pi i \al\al'}e^{2\pi i \al\mu}\sum_{n=0}^{\infty}e^{4\pi i n b\al}\mathcal{M}^n_{\al\al'}(\mu),\quad \cm^0_{\al\al'}(\mu)=1
\end{eqnarray}
Substituting this ansatz into the equation we find a simple recurrence relation between the corrections
\begin{multline}
\left(e^{2\pi ib\al'}e^{2\pi i b^2 n}e^{\f{b}2\p_\al}+e^{-2\pi ib\al'}e^{-2\pi i b^2 n}e^{-\f{b}2\p_\al}-2\cos{2\pi b\al'}\right)\mathcal{M}^{n}_{\al\al'}(\mu)=\\\left(e^{2\pi ib\mu}e^{2\pi ib\al'}e^{2\pi i b^2 (n-1)}e^{\f{b}2\p_\al}+e^{-2\pi ib\mu}e^{-2\pi ib\al'}e^{-2\pi i b^2 (n-1)}e^{-\f{b}2\p_\al}-2\cos{2\pi b\al'}\right)\mathcal{M}^{n-1}_{\al\al'}(\mu)\label{recurrence relation}
\end{multline}
One possible solution to this recurrence relation is given by $\al$-independent functions $\cm^n_{\al\al'}(\mu)=\cm^n_{\al'}(\mu)$. In this case shift operators act trivially and the recurrence relation is reduced to
\begin{multline}
\left(e^{2\pi ib\al'}e^{2\pi i b^2 n}+e^{-2\pi ib\al'}e^{-2\pi i b^2 n}-2\cos{2\pi b\al'}\right)\mathcal{M}^{n}_{\al'}(\mu)=\\\left(e^{2\pi ib\mu}e^{2\pi ib\al'}e^{2\pi i b^2 (n-1)}+e^{-2\pi ib\mu}e^{-2\pi ib\al'}e^{-2\pi i b^2 (n-1)}-2\cos{2\pi b\al'}\right)\mathcal{M}^{n-1}_{\al'}(\mu)
\end{multline}
Making use of the elementary trigonometric identities we can write the solution as follows
\begin{eqnarray}
\mathcal{M}^n_{\al'}(\mu)=\prod_{k=1}^{n}\f{\sin{\pi b(2\al'+\mu+b(k-1))}\sin{\pi b(\mu+b(k-1))}}{\sin{\pi b(2\al'+bk)}\sin{\pi b^2k}}\label{ratio of sines}
\end{eqnarray}
Given a function $\mathcal{M}^{n-1}_{\al\al'}(\mu)$ equation \eqref{recurrence relation} can be viewed as a second-order difference equation on $\mathcal{M}^{n}_{\al\al'}(\mu)$. It therefore admits two solutions. The first if $\al$-independent, as we just found. The second is, schematically
\begin{eqnarray}
\mathcal{M}^n_{\al\al'}(\mu)\propto e^{-8\pi i\al\al'}e^{-8\pi i n b\al} \mathcal{M}^n_{\al'}(\mu)
\end{eqnarray}
Appearance of the term with $e^{-8\pi i n b\al}$ means that we can not use such correction in the expansion \eqref{trans-series ansatz} which is performed in powers of $e^{4\pi i b\al}$. 

Therefore, formula \eqref{ratio of sines} is the only solution to the recurrence relation compatible with the expansion \eqref{trans-series ansatz}. One can also explain why the other solution is irrelevant by the following simple argument. Initial equation is of the second order and therefore admits exactly two independent solutions. We have chosen one of the branches with the choice of the zeroth-order correction \eqref{zeroth order choice}. After that, all the remaining corrections must be determined uniquely  and therefore the recurrence relation must admit a single legitimate solution.

Hence, we write the solution to equation \eqref{maineq a} as
\begin{eqnarray}
\mathcal{M}_{\al\al'}(\mu)=e^{4\pi i \al\al'}e^{2\pi i \al\mu}\sum_{n=0}^{\infty}e^{4\pi i n b\al} \mathcal{M}^n_{\al'}(n) \times f(\al,\al',\mu) \label{a-solution}
\end{eqnarray}
with coefficients given by equation \eqref{ratio of sines}. Here $f$ is any $b/2$-periodic in $\al$ function. We can partially fix it by imposing that expression \eqref{a-solution} also satisfies the other two equations \eqref{maineq a'},\eqref{maineq mu}. It appears, that the following function solves both, equation \eqref{maineq a} and \eqref{maineq a'}
\begin{multline}
\mathcal{M}_{\al\al'}(\mu)=e^{4\pi i \al\al'}e^{2\pi i\al\mu}\sin{2\pi b\al'}\f{S_b(2\al'+\mu)}{S_b(2\al'+Q)}\times\\\sum_{n=0}^{\infty}e^{4\pi i n b\al} \prod_{k=1}^{n}\f{\sin{\pi b(2\al'+\mu+b(k-1))}\sin{\pi b(\mu+b(k-1))}}{\sin{\pi b(2\al'+bk)}\sin{\pi b^2k}} \times g(\al,\al',\mu) \label{aa'-solution}
\end{multline}
with arbitrary function $g(\al,\al',\mu)$ which is $b/2$-periodic in both $\al$ and $\al'$. To show this we use the following series representation for the double sine function
\begin{eqnarray}
\log{S_{b}(z)}=-\f{i\pi}{2}\left(z^2-Qz+\f{Q^2+1}{6}\right)-\sum_{n=1}^{\infty}\f1n\f{e^{2\pi i nb z}}{e^{2\pi i nb^2 }-1}-\sum_{n=1}^{\infty}\f1n\f{e^{2\pi i nb^{-1} z}}{e^{2\pi i nb^{-2} }-1}
\end{eqnarray}
Let us also introduce notation
\begin{eqnarray}
\log S_0(z|Q)=-\f{i\pi}{2}\left(z^2-Qz+\f{Q^2+1}{6}\right)\\
\log \widetilde{S}(z|b)=-\sum_{n=1}^{\infty}\f1n\f{e^{2\pi i nb z}}{e^{2\pi i nb^2 }-1}
\end{eqnarray}
so that the double sine function is split into three factors
\begin{eqnarray}
S_b(z)=S_0(z|Q)\widetilde{S}(z|b)\widetilde{S}(z|b^{-1})
\end{eqnarray}
First, note that $\widetilde{S}(z|b^{-1})$ is $b/2$-periodic and therefore can be absorbed into $g(\al,\al',\mu)$. Next, straightforwardly \begin{eqnarray}
\f{S_0(2\al'+\mu)}{S_0(2\al'+Q)}\sim \f{S_0(2\al'+\mu)}{S_0(2\al'+b)}=e^{2\pi i (b-\mu) \al'}e^{i\pi\mu(b-\mu)/2}
\end{eqnarray}
and we have omitted $b/2$-periodic function. Finally,
\begin{multline}
\f{\widetilde{S}(2\al'+\mu|b)}{\widetilde{S}(2\al'|b)}\prod_{k=1}^{n}\f{\sin{\pi b(2\al'+\mu+(k-1)b)}\sin{\pi b(\mu+(k-1)b)}}{\sin{\pi b(2\al'+kb)}\sin{\pi b^2k}}=\\\sum_{m=0}^{\infty} e^{4\pi i mb\al'}e^{2\pi i b^2 n m}\prod_{k=1}^{n}\f{e^{2\pi i kb^2}-e^{2\pi i b(b-\mu)}}{e^{2\pi i kb^2}-1}\prod_{l=1}^{m}\f{e^{2\pi i lb^2}-e^{2\pi i b\mu}}{e^{2\pi i lb^2}-1}
\end{multline} 
This formula is easily verified on a computer in any desired number of orders. Though the analytical proof should not be hard, we do not attempt to give it. This formula serves as a simplified version of the correction coefficients \eqref{ratio of sines}. It also is manifestly symmetric w.r.t. to the permutation $\al\leftrightarrow\al'$ supplemented with the change $\mu\to b-\mu$. Therefore, under these transformations function \eqref{aa'-solution} behaves as
\begin{eqnarray}
\mathcal{M}_{\al'\al}(b-\mu)\sim \mathcal{M}_{\al\al'}(\mu) \f{\sin{2\pi b\al}}{\sin{2\pi b \al'}}\label{eq exhange}
\end{eqnarray}
where again, we did not account for periodic terms.
As can be seen from comparison of \eqref{maineq a} and \eqref{maineq a'} transformation \eqref{eq exhange} converts solution of one of these equations to a solution of the other. Therefore, function \eqref{aa'-solution}, which is invariant under this transformation up to $b/2$-periodic terms, solves both of them. Moreover, it is not hard to verify that the same function solves the last equation \eqref{maineq mu}. In other words, we do not have to include any non-trivial dependence on $\mu$ in function $g(\al,\al',\mu)$. 

So far we have seen that function
\begin{multline}
\mathcal{M}_{\al\al'}(\mu)=e^{4\pi i \al\al'}e^{2\pi i (\mu \al+(b-\mu)\al')}e^{\f{i\pi}2\mu(b-\mu)}\sin{2\pi b\al'}\times\\\sum_{n,m=0}^{\infty}e^{4\pi i nb\al} e^{4\pi i mb\al'}e^{2\pi i b^2 n m}\prod_{k=1}^{n}\f{e^{2\pi i kb^2}-e^{2\pi i b(b-\mu)}}{e^{2\pi i kb^2}-1}\prod_{l=1}^{m}\f{e^{2\pi i lb^2}-e^{2\pi i b\mu}}{e^{2\pi i lb^2}-1}
\end{multline}
 solves each of equations \eqref{maineq a},\eqref{maineq a'},\eqref{maineq mu}. It remains to impose symmetry w.r.t. transformation $b\to b^{-1}$ and to render the function even in $\al'$. The procedure is straightforward and leads to formula \eqref{mk}. Overall constant normalization factor (which appears to be  $2^{3/2}$) is, of course, not quite important, but if needed can be determined from condition $M^2=1$ and is most easily  verified for $\mu=0$.

\section{Series expansion from the integral representation\label{sec series from integral}}
In this section we derive the series expansion \eqref{mk} directly from the integral representation \eqref{MK by PT}. As turns out, the series expansion corresponds to the sum over the residues of the integral. Function $S_b(z)$ is meromorphic with simple poles and zeros located at
\begin{align}
&\text{zeroes}: z=nb+mb^{-1},\quad n,m\ge 1\nn\\
&\text{poles}: z=-nb-mb^{-1},\quad n,m\ge 0
\end{align} 
Therefore, the integrand of \eqref{MK by PT} has 
simple poles at the points
\begin{align}
&\xi_{II}=\al'+Q-\mu/2-nb-mb^{-1},\quad n,m\ge 1 &\qquad &\xi_{I}=\al'+\mu/2+nb+mb^{-1},\quad n,m\ge0&\nn\\&\xi_{III}=-\al'-\mu/2-nb-mb^{-1},\quad n,m\ge 0& \qquad &\xi_{IV}=-\al'-Q+\mu/2+nb+mb^{-1},\quad n,m\ge1&\label{integrand poles}
\end{align}
We can depict them in the complex $\xi$-plane
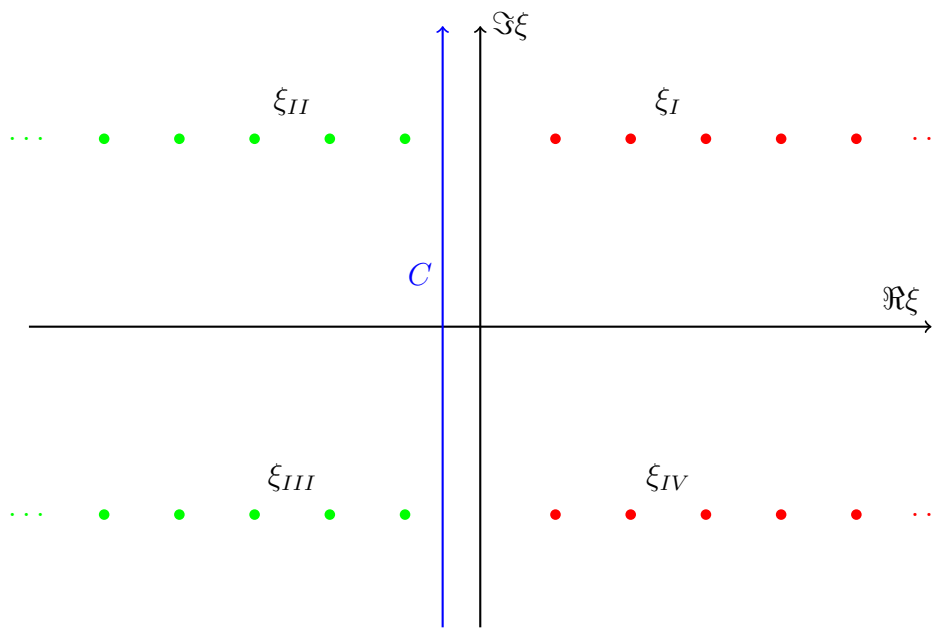
\begin{figure}[H]
\centering
\begin{tikzpicture}
\draw[thick,->] (-6,0) -- (6,0) node[anchor = south east] {$\Re\xi$};
\draw[thick,->] (0,-4) -- (0,4) node[anchor = west] {$\Im \xi$};
\draw[thick,blue,->] (-0.5,-4) -- (-0.5,4);
\node[blue] at (-0.8,0.7) {$C$};
\foreach \x in {1,2,3,4,5}
	\fill[red] (\x,2.5) circle (2pt);
\node[red] at (6, 2.5) {$\dots$};
\foreach \x in {1,2,3,4,5}	 
	\fill[red] (\x,-2.5) circle (2pt);
\node[red] at (6, -2.5) {$\dots$};
\foreach \x in {-1,-2,-3,-4,-5}	 
	\fill[green] (\x,2.5) circle (2pt);
\node[green] at (-6, 2.5) {$\dots$};
\foreach \x in {-1,-2,-3,-4,-5}	 
	\fill[green] (\x,-2.5) circle (2pt);
\node[green] at (-6, -2.5) {$\dots$};
\node at (2.5, 3) {$\xi_I$};
\node at (-2.5, 3) {$\xi_{II}$};
\node at (-2.5, -2) {$\xi_{III}$};
\node at (2.5, -2) {$\xi_{IV}$};
\end{tikzpicture}
\end{figure}
\noindent Possible integration contour $C$ in \eqref{MK by PT} is shown by the blue line. Assume now that $\al\in i\mathbb{R}_{+}$ and $b>0$. Then exponent $e^{-4\pi i \al\xi}$ in \eqref{MK by PT} decays for $\Re \xi<0$ and the integral can be represented as the sum over the residues collected at $z=\xi_{II}$ and $z=\xi_{III}$.

Note that this figure is a little schematic. In each of the four families poles are not located equidistantly; for complex values of $b$ one would have wedges instead of half-lines; in some circumstances the four groups are not sharply separated and the contour of integration can not be chosen to run along the imaginary axes (we have shifted the contour at the figure from imaginary axis for the clarity of the picture). However, no matter what the deformation is necessary to correct the figure in any particular case, the result remains simple: we can compute integral \eqref{MK by PT} accounting for residues $\xi_{II}$ and $\xi_{III}$.

Denote the integrand of \eqref{MK by PT} by $\mathcal{I}$, then
\begin{eqnarray}
Res_{\xi_{III}}\mathcal{I}=\f{S_b(2\al'+\mu+nb+mb^{-1})S_b(\mu+nb+mb^{-1})}{S_b(2\al'+Q+nb+mb^{-1})Res^{-1}S_b(-nb-mb^{-1})}e^{4\pi i\al\al'}e^{2\pi i\al\mu}e^{4\pi i\al(nb+mb^{-1})}
\end{eqnarray}
From definition \eqref{integrand poles} one sees that the poles corresponding to $\xi_{II}$ describe the same set as $\xi_{III}$ but with $\al'$ replaced by $-\al'$ (recall that $Q=b+b^{-1}$). Also, the integrand is $\al'$-even function as can be seen from property \eqref{Sb inverse}. As a consequence the residues of the integrand at $\xi_{II}$ are exactly the same as at $\xi_{III}$ with $\al'$ replaced by $-\al'$. Thus we can only consider the sum over $\xi_{III}$ residues and then symmetrize the result to get an even function of $\al'$.

Using property \eqref{Sb difference relation} and explicit form of residues \eqref{residues}  we can rewrite the obtained expression in the following way
\begin{multline}
Res_{\xi_{III}}\mathcal{I} =e^{4\pi i\al\al'}e^{2\pi i\al\mu}e^{4\pi i\al(nb+mb^{-1})} S_b(\mu)\f{S_b(2\al'+\mu)}{S_b(2\al'+Q)}\times\\\prod_{k=1}^{n}\f{\sin{\pi b(2\al'+\mu+(k-1)b)}\sin{\pi b(\mu+(k-1)b)}}{\sin{\pi b(2\al'+kb)}\sin{\pi b^2k}}\times\\\prod_{l=1}^{m}\f{\sin{\pi b^{-1}(2\al'+\mu+(l-1)b^{-1})}\sin{\pi b^{-1}(\mu+(l-1)b^{-1})}}{\sin{\pi b^{-1}(2\al'+lb^{-1})}\sin{\pi b^{-2}l}}\label{temp residue}
\end{multline}
This formula is almost the same as \eqref{aa'-solution}, from which representation \eqref{mk} is derived. The difference disappears completely if we take into account factor $S_b^{-1}(\mu)\sin{2\pi b\al'}$ from formula \eqref{MK by PT}.
 Hence we see, that summing up all the contributions from $\xi_{III}$ and $\xi_{II}$ groups of poles reproduces formula \eqref{aa'-solution} while automatically choosing the $b^{-1/2}$-remainder $g(\al,\al',\mu)$ in the right way. We conclude that formula \eqref{mk} indeed serves as a series representation to the original integral expression \eqref{MK by PT}. 

Note  that for $\al\in -i\mathbb{R}_+$ we can enclose the integration contour in \eqref{MK by PT} in the right half-plane collecting $\xi_{I}$ and $\xi_{IV}$ residues. As is seen from \eqref{integrand poles} and the symmetry of the integrand the result will be exactly the same up to replacement $\al\to-\al$. This simply means that $\cm_{\al\al'}(\mu)$ is even w.r.t. to $\al$, as it should be. Thus, representation \eqref{mk} can be trivially continued to $\al\in -i\mathbb{R}_+$.
\section*{Acknowledgements}
The author is grateful to all the people he has engaged in discussions on the subject and especially to Alexei Morozov and Andrei Mironov for their great assistance. The work is partly supported by grant NSh-1500.2014.2, by RFBR 13-02-00457-a, by RFBR 14-02-31372 mol\_a, by ${15-51-52031}$-NSC-a, and by 15-51-50041-YaF.
\appendix
\section{Special functions\label{app special functions}}
Double gamma function $\g_b(z)$ can be defined by means of the integral representation
\begin{eqnarray}
\log \g_b(z)=\int_0^\infty \frac{dt}{t}\left(\frac{e^{-z t}-e^{-Q t/2}}{\left(1-e^{-b t}\right)
   \left(1-e^{-b^{-1}t}\right)}-\frac{(Q-2 z)^2}{8 e^t}-\frac{Q-2
   z}{2t}\right),\qquad Q=b+b^{-1}\label{double gamma integral definition}
\end{eqnarray}
For our purposes the main property of this function is the following difference relation
\begin{eqnarray}
\g_b(z+b)=\f{\s{2\pi}\,b^{bz-1/2}}{\g(bz)}\g_b(z)\label{double gamma difference relation}
\end{eqnarray}
Function $S_b(z)$ can be defined by
\begin{eqnarray}
S_b(z)=\f{\g_b(z)}{\g_b(Q-z)}\label{double sine definition}
\end{eqnarray}
From the definition one derives
\begin{eqnarray}
S_b(Q-z)=\f1{S_b(z)}\label{Sb inverse}\\
S_b(z+b)=2\sin{\pi b z}\, S_b(z)\label{Sb difference relation}
\end{eqnarray}
Note that $\Gamma_{b}(z)=\Gamma_{b^{-1}}(z),\,\, S_{b}(z)=S_{b^{-1}}(z)$, property often called self-duality. 

Double sine function has poles at points $z=-nb-mb^{-1} (n,m\ge 0)$ and zeros at the points $z=nb+mb^{-1} (n,m\ge 1)$. The corresponding residues are
\begin{align}
&Res\, S_b(-nb-mb^{-1})=\f1{2\pi}\f{(-1)^{nm+n+m}}{\prod_{k=1}^n 2\sin{\pi kb^2\prod_{l=1}^m 2\sin{\pi lb^{-2}}}}\nn\\
&Res\, S^{-1}_b(nb+mb^{-1})= \f1{2\pi}\f{(-1)^{nm}}{\prod_{k=1}^{n-1} 2\sin{\pi kb^2\prod_{l=1}^{m-1} 2\sin{\pi lb^{-2}}}} \label{residues}
\end{align}

\section{Derivation of the difference equations on the morular kernel \label{app derivation of equations}}
In this Appendix we briefly present a derivation of equations \eqref{maineq a}, \eqref{maineq a'}, \eqref{maineq mu}. Recall that in order to obtain the pentagon identity one needs to consider transformation the properties of the five-point conformal blocks. Likewise, in order to derive equations on the toric modular kernel we have to consider not a one-point toric conformal block, but a two-point block
\begin{figure}[H]
\centering
\begin{tikzpicture}
\draw[line width=2pt] (3,0)--(4,0);
\draw[line width=2pt] (2,1) node[above, left] {$m_1$} -- (3,0);
\draw[line width=2pt] (2,-1) node[above, left] {$m_2$} -- (3,0) node[above right] {$\mu$};
\draw[line width=2pt] (4.7,0) circle (20 pt);
\node[left] at (1,0) {$B_a(
m_1, m_2;\mu|\tau)
= $};
\node[above] at (4.7, 22 pt) {$a$};
\end{tikzpicture}
\end{figure}
\noindent Operators corresponding to the legs $m_1,m_2$ are inserted close to each other somewhere on the torus. Conformal block also depends on their relative position, but we will not denote this dependence in a manifest way. 

When one of the legs, say $m_2$ travels a closed path around the torus the conformal block acquires monodromy. We can express this monodromy via the \textit{the fusion matrices} by representing a closed path as a number of analytic continuations each performed by the corresponding operation on the four-point block. Subsequent manipulations are close to those used to obtain the Verlinde \cite{Verlinde:1988sn} formula and were extensively applied in papers \cite{Drukker:2009id, Alday:2009fs} to compute Wilson/'t Hooft loops in AGT dual gauge theories.

 Assume that a closed loop in our pictorial representation of the two-point block is the $B$-cycle of the torus. Then, the series of moves that represent transport of $m_2$ operator along the $A$ cycle is 
 
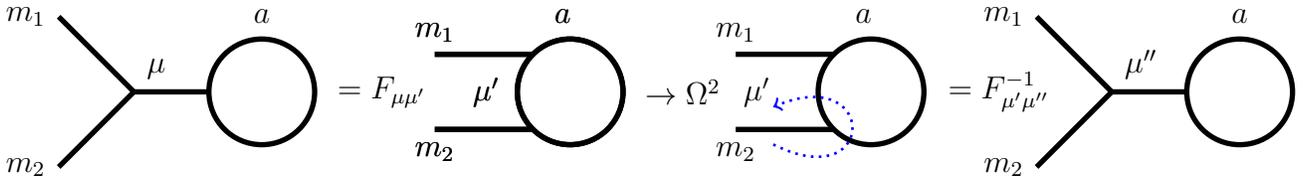
\begin{figure}[H]
\centering
\begin{tikzpicture}

\draw[line width=2pt] (3,0)--(4,0);
\draw[line width=2pt] (2,1) node[above, left] {$m_1$} -- (3,0);
\draw[line width=2pt] (2,-1) node[above, left] {$m_2$} -- (3,0) node[above right] {$\mu$};
\draw[line width=2pt] (4.7,0) circle (20 pt);
\node[above] at (4.7, 22 pt) {$a$};

\node at (6.3,0) {$= F_{\mu\mu'}$};

\draw[line width=2pt] (7,0.5) node[above] {$m_1$} -- (8.3,0.5);
\draw[line width=2pt] (7,-0.5) node[below] {$m_2$} -- (8.3,-0.5);
\draw[line width=2pt] (8.8,0) circle (20 pt);
\node[left] at (8,0) {$\mu'$};
\node[above] at (8.7, 22 pt) {$a$};

\node at  (10.3,0) {$\to\Omega^2$};

\draw[line width=2pt] (11,0.5) node[above] {$m_1$} -- (12.3,0.5);
\draw[line width=2pt] (11,-0.5) node[below] {$m_2$} -- (12.3,-0.5);
\draw[line width=2pt] (12.8,0) circle (20 pt);
\node[left] at (11.6,0) {$\mu'$};
\node[above] at (12.7, 22 pt) {$a$};
\draw[blue,line width=1pt,->,dotted] (11.5, -0.7) .. controls (12.9,-1.4) and (12.9,0.4) .. (11.5, -0.2);

\node at (14.5,0) {$= F^{-1}_{\mu'\mu''}$};

\draw[line width=2pt] (7,0.5) node[above] {$m_1$} -- (8.3,0.5);
\draw[line width=2pt] (7,-0.5) node[below] {$m_2$} -- (8.3,-0.5);
\draw[line width=2pt] (8.8,0) circle (20 pt);
\node[left] at (8,0) {$\mu'$};
\node[above] at (8.7, 22 pt) {$a$};

\draw[line width=2pt] (16,0)--(17,0);
\draw[line width=2pt] (15,1) node[above, left] {$m_1$} -- (16,0);
\draw[line width=2pt] (15,-1) node[above, left] {$m_2$} -- (16,0) node[above right] {$\mu''$};
\draw[line width=2pt] (17.7,0) circle (20 pt);
\node[above] at (17.7, 22 pt) {$a$};

\node at (1,3) {};
\node at (1,-3) {};
\end{tikzpicture}
\end{figure}
\noindent Therefore, we have the following expression for the conformal block continued along $A$-cycle
\begin{align}
&A\circ B_a(
m_1, m_2;\mu|\tau)=\sum_{\mu',\mu''}F_{\mu\mu'}\begin{bmatrix}
m_1&a\\m_2&a
\end{bmatrix}\Omega^2(\mu';m_2,a)F^{-1}_{\mu'\mu''}\begin{bmatrix}
m_1&a\\m_2&a
\end{bmatrix}B_{a}(
m_1, m_2;\mu''|\tau)
\end{align}
In words: the first move uncouples two external legs; the second move transports the $m_2$ leg along $A$-cycle (denoted by dotted arrow); the third move fuses two external operators back together. Quantities $F_{aa'}$ are the same fusion matrices that relate s- and t-channel spheric conformal blocks while $\Omega(a_1, a_2, a_3)$ is the phase factor representing monodromy of permutation of the two legs in the conformal block which are 'close' to each other. It can be read off from the OPE  asymptotic and equals simply
\begin{eqnarray}
\Omega(a_1; a_2, a_3)=e^{i\pi (\Delta(a_1)-\Delta(a_2)-\Delta(a_3))}
\end{eqnarray}

Similarly, $B$-cycle monodromy is represented as
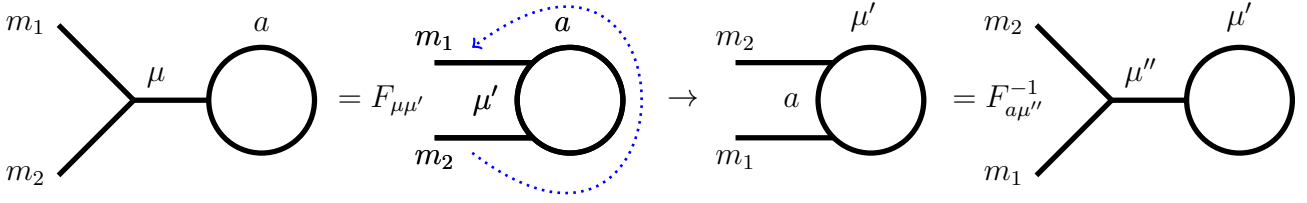
\begin{figure}[H]
\centering
\begin{tikzpicture}

\draw[line width=2pt] (3,0)--(4,0);
\draw[line width=2pt] (2,1) node[above, left] {$m_1$} -- (3,0);
\draw[line width=2pt] (2,-1) node[above, left] {$m_2$} -- (3,0) node[above right] {$\mu$};
\draw[line width=2pt] (4.7,0) circle (20 pt);
\node[above] at (4.7, 22 pt) {$a$};

\node at (6.3,0) {$= F_{\mu\mu'}$};

\draw[line width=2pt] (7,0.5) node[above] {$m_1$} -- (8.3,0.5);
\draw[line width=2pt] (7,-0.5) node[below] {$m_2$} -- (8.3,-0.5);
\draw[line width=2pt] (8.8,0) circle (20 pt);
\node[left] at (8,0) {$\mu'$};
\node[above] at (8.7, 22 pt) {$a$};
\draw[blue,line width=1pt,->,dotted] (7.5, -0.7) .. controls (10.5,-3) and (10.5, 3) .. (7.5, 0.7);

\node at  (10.3,0) {$\to$};

\draw[line width=2pt] (11,0.5) node[above] {$m_2$} -- (12.3,0.5);
\draw[line width=2pt] (11,-0.5) node[below] {$m_1$} -- (12.3,-0.5);
\draw[line width=2pt] (12.8,0) circle (20 pt);
\node[left] at (12.0,0) {$a$};
\node[above] at (12.7, 22 pt) {$\mu'$};

\node at (14.5,0) {$= F^{-1}_{a\mu''}$};

\draw[line width=2pt] (7,0.5) node[above] {$m_1$} -- (8.3,0.5);
\draw[line width=2pt] (7,-0.5) node[below] {$m_2$} -- (8.3,-0.5);
\draw[line width=2pt] (8.8,0) circle (20 pt);
\node[left] at (8,0) {$\mu'$};
\node[above] at (8.7, 22 pt) {$a$};

\draw[line width=2pt] (16,0)--(17,0);
\draw[line width=2pt] (15,1) node[above, left] {$m_2$} -- (16,0);
\draw[line width=2pt] (15,-1) node[above, left] {$m_1$} -- (16,0) node[above right] {$\mu''$};
\draw[line width=2pt] (17.7,0) circle (20 pt);
\node[above] at (17.7, 22 pt) {$\mu'$};

\end{tikzpicture}
\end{figure}
\noindent The first and the last moves here are the same as for the A-cycle. However, in contrast to the A-cycle transport along the direction of the B-cycle does not introduce a phase factor but simply permutes the two intermediate dimensions $a\leftrightarrow\mu'$. Quantitatively
\begin{align}
&B\circ B_a(
m_1, m_2;\mu|\tau)=\sum_{\mu',\mu''}F_{\mu\mu'}\begin{bmatrix}
m_1&a\\m_2&a
\end{bmatrix}F^{-1}_{a\mu''}\begin{bmatrix}
m_2&\mu'\\m_1&\mu'
\end{bmatrix}B_{\mu'}(
m_2, m_1;\mu''|\tau)
\end{align}
\nocite{apsrev41Control}
Now, since the modular S-transformation permutes A and B cycles we have the following consistency condition
\begin{eqnarray}
S\circ A=B\circ S
\end{eqnarray}
Spelled out explicitly it reads
\begin{multline}
M_{aa'}(\mu)\sum_{\mu''}F_{\mu\mu''}\begin{bmatrix}
m_1&a'\\m_2&a'
\end{bmatrix}\Omega^2(\mu'';m_2,a')F^{-1}_{\mu''\mu'}\begin{bmatrix}
m_1&a'\\m_2&a'
\end{bmatrix}=\\\sum_{a''}F_{\mu a''}\begin{bmatrix}
m_1&a\\m_2&a
\end{bmatrix}F^{-1}_{a\mu'}\begin{bmatrix}
m_2&a''\\m_1&a''
\end{bmatrix} M_{a''a'}(\mu')\label{gen eq}
\end{multline}
This is the analog of the pentagon identity intertwining spheric and toric transformations. 

We can turn this equation into a second-order difference equation along the lines described in subsection \ref{pentagon section}. Namely, set momentum $m_2$ to a degenerate value $m_2=-b/2$. Due to fusion rules \eqref{fusion rules} for $m_2=-b/2$ we have the following selection rules on momenta entering equation \eqref{gen eq}
\begin{eqnarray}
\mu=m_1+s_1b/2,\quad \mu''=m_1+s_2b/2,\quad \mu'=a'+s_3b/2,\quad a''=a+s_4b/2
\end{eqnarray}  
where all quantities $s_1,s_2,s_3,s_4$ are either $+$ or $-$. Denote
\begin{eqnarray}
F_{a=a_1+s_1b/2,\,a'=a_3+s_2b/2}\begin{bmatrix}
-b/2&a_3\\a_1&a_4
\end{bmatrix}=F_{s_1,s_2}\begin{bmatrix}
-b/2&a_3\\a_1&a_4
\end{bmatrix}
\end{eqnarray}
Then, equation \eqref{gen eq} for $m_2=-b/2$ can be rewritten as
\begin{multline}
M_{aa'}(\mu)\sum_{s_2=\pm} F_{s_1,s_2}\begin{bmatrix}
\mu-s_1b/2& a'\\-b/2&a'
\end{bmatrix}\Omega^2(a'+s_2b/2;-b/2,a')F^{-1}_{s_2,s_3}\begin{bmatrix}
\mu-s_1b/2& a'\\-b/2&a'
\end{bmatrix}=\\\sum_{s_4=\pm}F_{s_1,s_4}\begin{bmatrix}
\mu-s_1b/2& a\\-b/2&a
\end{bmatrix} F^{-1}_{-s_4,s_3}\begin{bmatrix}
\-b/2& a+s_4b/2\\\mu-s_1b/2&a+s_4b/2
\end{bmatrix}M_{a+s_4b/2,a'}(\mu+(s_3-s_1)b/2)\label{gen diff eq}
\end{multline}
This is indeed a difference equation on the modular kernel $M_{aa'}(\mu)$. Note that for $s_1=s_3$ only shifts in the internal momentum $a$ are presented in the equation, while for $s_1\neq s_3$ we also have shifts in the external momentum $\mu$.

  Moreover, all of the fusion matrices entering equation \eqref{gen diff eq} are known explicitly \eqref{degenerate fusion matrices}. In we renormalize vertices according to \eqref{toric normalization} these matrices greatly simplify. Namely, for generic values of parameters the transformation is 

\begin{eqnarray}
\cf_{aa'}\begin{bmatrix}
a_2&a_3\\a_1&a_4
\end{bmatrix}=\f{V(a_1,a_2;a)V(a,a_3;a_4)}{V(a_3,a_2;a')V(a', a_1;a_4)} F_{aa'}\begin{bmatrix}
a_2&a_3\\a_1&a_4
\end{bmatrix}
\end{eqnarray}
If we now set $m_2=-b/2$ then, according to the fusion rules \eqref{fusion rules} 
\begin{eqnarray}
a=a_1+ s_1 b/2,\quad a'=a_3+ s_2b/2
\end{eqnarray}
with $s_1,s_2=\pm$. For these special choices of parameters the ratios of the double gamma functions in the renormalization factor reduce to the ordinary gamma functions which in turn almost cancel with those in the standard normalization \eqref{degenerate fusion matrices}. The result is
\begin{eqnarray}
 \cf_{s_1,s_2}\begin{bmatrix}-b/2 & a_3\\ a_1 & a_4\end{bmatrix} =s_1\f{\sin{\pi b(a_4+s_1 a_3-s_2 a_1 -(1+s_1-s_2)b/2))}}{\sin{\pi b(2a_3-b)}}\label{Fa2}
 \end{eqnarray} 
We can obtain four equations from \eqref{gen diff eq} corresponding to each choice of $s_1,s_3=\pm$.
Equations for $s_1=s_3=1$ and $s_1=-s_3=1$ are
\begin{align}
 \left(\f{\sin{\pi b(2a-b+\mu)}}{\sin{\pi b(2a-b)}}e^{\f{b}{2}\p_a}+\f{\sin{\pi b(2a-b-\mu)}}{\sin{\pi b(2a-b)}}e^{-\f{b}{2}\p_a}\right)\cm_{aa'}(\mu)&= -2\cos{\pi b(2a'-b)}\cm_{aa'}(\mu)\nn\\
 \frac{1}{\sin{\pi b(2a-b)}}\left(e^{\f{b}{2}\p_a}-e^{-\f{b}{2}\p_a}\right)\cm_{aa'}(\mu-b)&=2\cm_{aa'}(\mu)
 \end{align} 
 Equations with $s_1=s_3=-1$ and $s_1=-s_3=-1$ are equivalent to the above equations. Upon redefinition
 \begin{eqnarray}
 a=\al+Q/2,\quad a'=\al'+Q/2
 \end{eqnarray}
 these equations become \eqref{maineq a} and \eqref{maineq mu}. Equation \eqref{maineq a'} can be derived from condition
 \begin{eqnarray}
 \int\, d\al' \cm_{\al\al'}(\mu) \cm_{\al'\al''}(\mu)=\delta(\al-\al'')
 \end{eqnarray}
Namely,
\begin{multline}
\int d\al' \cm_{\al\al'}(\mu)\cm_{\al'\al''}(\mu)=\\\f1{2\cos{2\pi b\al''}}\int d\al' \cm_{\al\al'}(\mu)\left(\f{\sin{\pi b(2\al'+\mu)}}{\sin{2\pi b\al'}}e^{\f{b}2\p_{\al'}}+\f{\sin{\pi b(2\al'-\mu)}}{\sin{2\pi b\al'}}e^{-\f{b}2\p_{\al'}}\right)\cm_{\al'\al''}(\mu)=\\\f1{2\cos{2\pi b\al''}}\int d\al' \left[\left(e^{-\f{b}2\p_{\al'}}\f{\sin{\pi b(2\al'+\mu)}}{\sin{2\pi b\al'}}+e^{\f{b}2\p_{\al'}}\f{\sin{\pi b(2\al'-\mu)}}{\sin{2\pi b\al'}}\right)\cm_{\al\al'}(\mu)\right]\cm_{\al'\al''}(\mu)=\\\delta(\al-\al'')
\end{multline}
Inverting the last equality (integrating over $\al''$ with the measure $\cm_{\al''\al'}(\mu)$) we recover equation \eqref{maineq a'}. This completes the derivation of our main equations \eqref{maineq a},\eqref{maineq a'},\eqref{maineq mu}.
\nocite{apsrev41Control}
\bibliographystyle{utcaps_edited}
\bibliography{bibfile,revtex-custom}

\providecommand{\href}[2]{#2}\begingroup\raggedright\begin{thebibliography}{10}

\bibitem{GMMpert}
D.~Galakhov, A.~Mironov, and A.~Morozov, ``{S-duality as a beta-deformed
  Fourier transform},'' \href{http://dx.doi.org/10.1007/JHEP08(2012)067}{{\em
  JHEP} {\bfseries 1208} (2012) 067},
\href{http://arxiv.org/abs/1205.4998}{{\ttfamily arXiv:1205.4998 [hep-th]}}.

\bibitem{Billo:2013fi}
M.~Billo, M.~Frau, L.~Gallot, A.~Lerda, and I.~Pesando, ``{Deformed N=2
  theories, generalized recursion relations and S-duality},''
  \href{http://dx.doi.org/10.1007/JHEP04(2013)039}{{\em JHEP} {\bfseries 1304}
  (2013) 039},
\href{http://arxiv.org/abs/1302.0686}{{\ttfamily arXiv:1302.0686 [hep-th]}}.

\bibitem{Lerda}
M.~Billo, M.~Frau, L.~Gallot, A.~Lerda, and I.~Pesando, ``{Modular anomaly
  equation, heat kernel and S-duality in $N=2$ theories},''
  \href{http://dx.doi.org/10.1007/JHEP11(2013)123}{{\em JHEP} {\bfseries 1311}
  (2013) 123},
\href{http://arxiv.org/abs/1307.6648}{{\ttfamily arXiv:1307.6648 [hep-th]}}.

\bibitem{Nemkov1}
N.~Nemkov, ``{S-duality as Fourier transform for arbitrary
  $\epsilon_1,\epsilon_2$},''
  \href{http://dx.doi.org/10.1088/1751-8113/47/10/105401}{{\em J. Phys. A:
  Math. Theor.} {\bfseries 47} (2014) 105401},
\href{http://arxiv.org/abs/1307.0773}{{\ttfamily arXiv:1307.0773 [hep-th]}}.

\bibitem{PT1}
B.~Ponsot and J.~Teschner, ``{Liouville bootstrap via harmonic analysis on a
  noncompact quantum group},''
\href{http://arxiv.org/abs/hep-th/9911110}{{\ttfamily arXiv:hep-th/9911110
  [hep-th]}}.

\bibitem{PT2}
B.~Ponsot and J.~Teschner, ``{Clebsch-Gordan and Racah-Wigner coefficients for
  a continuous series of representations of U(q)(sl(2,R))},''
  \href{http://dx.doi.org/10.1007/PL00005590}{{\em Commun.Math.Phys.}
  {\bfseries 224} (2001) 613--655},
\href{http://arxiv.org/abs/math/0007097}{{\ttfamily arXiv:math/0007097
  [math-qa]}}.

\bibitem{PT3}
J.~Teschner, ``{From Liouville theory to the quantum geometry of Riemann
  surfaces},''
\href{http://arxiv.org/abs/hep-th/0308031}{{\ttfamily arXiv:hep-th/0308031
  [hep-th]}}.

\bibitem{BPZ}
A.~Belavin, A.~Polyakov, and A.~Zamolodchikov, ``{Infinite Conformal Symmetry
  in Two-Dimensional Quantum Field Theory},''
\href{http://dx.doi.org/10.1016/0550-3213(84)90052-X}{{\em Nucl.Phys.}
  {\bfseries B241} (1984) 333--380}.

\bibitem{AGT}
L.~Alday, D.~Gaiotto, and Y.~Tachikawa, ``{Liouville Correlation Functions from
  Four-dimensional Gauge Theories},''
  \href{http://dx.doi.org/10.1007/s11005-010-0369-5}{{\em Lett.Math.Phys.}
  {\bfseries 91} (2010) 167--197},
\href{http://arxiv.org/abs/0906.3219}{{\ttfamily arXiv:0906.3219 [hep-th]}}.

\bibitem{Alba:2009ya}
V.~Alba and An. Morozov, ``{Check of AGT Relation for Conformal Blocks on
  Sphere},'' \href{http://dx.doi.org/10.1016/j.nuclphysb.2010.05.016}{{\em
  Nucl.Phys.} {\bfseries B840} (2010) 441--468},
\href{http://arxiv.org/abs/0912.2535}{{\ttfamily arXiv:0912.2535 [hep-th]}}.

\bibitem{Zamolodchikov:Rec_c}
Al. Zamolodchikov, ``{Conformal symmetry in two dimensions: an explicit
  recurrence formula for the conformal partial wave amplitude},''
\href{http://dx.doi.org/10.1007/BF01214585}{{\em Commun.Math.Phys.} {\bfseries
  96} (1984) 419--422}.

\bibitem{Zamolodchikov:Rec_a}
Al. Zamolodchikov, ``{Conformal symmetry in two-dimensional space: Recursion
  representation of conformal block},'' {\em Theor.Math.Phys.} {\bfseries 73}
  (1987) 1088--1093.

\bibitem{Itoyama:2014dya}
H.~Itoyama, A.~Mironov, and A.~Morozov, ``{Matching branches of
  non-perturbative conformal block at its singularity divisor},''
\href{http://arxiv.org/abs/1406.4750}{{\ttfamily arXiv:1406.4750 [hep-th]}}.

\bibitem{Poghossian}
R.~Poghossian, ``{Recursion relations in CFT and N=2 SYM theory},''
  \href{http://dx.doi.org/10.1088/1126-6708/2009/12/038}{{\em JHEP} {\bfseries
  0912} (2009) 038},
\href{http://arxiv.org/abs/0909.3412}{{\ttfamily arXiv:0909.3412 [hep-th]}}.

\bibitem{Hadasz:2009db}
L.~Hadasz, Z.~Jaskolski, and P.~Suchanek, ``{Recursive representation of the
  torus 1-point conformal block},''
  \href{http://dx.doi.org/10.1007/JHEP01(2010)063}{{\em JHEP} {\bfseries 1001}
  (2010) 063},
\href{http://arxiv.org/abs/0911.2353}{{\ttfamily arXiv:0911.2353 [hep-th]}}.

\bibitem{Hadasz:2009sw}
L.~Hadasz, Z.~Jaskolski, and P.~Suchanek, ``{Modular bootstrap in Liouville
  field theory},'' \href{http://dx.doi.org/10.1016/j.physletb.2010.01.036}{{\em
  Phys.Lett.} {\bfseries B685} (2010) 79--85},
\href{http://arxiv.org/abs/0911.4296}{{\ttfamily arXiv:0911.4296 [hep-th]}}.

\bibitem{Wyllard:2009hg}
N.~Wyllard, ``{A(N-1) conformal Toda field theory correlation functions from
  conformal N = 2 SU(N) quiver gauge theories},''
  \href{http://dx.doi.org/10.1088/1126-6708/2009/11/002}{{\em JHEP} {\bfseries
  0911} (2009) 002},
\href{http://arxiv.org/abs/0907.2189}{{\ttfamily arXiv:0907.2189 [hep-th]}}.

\bibitem{Mironov:2009qt}
A.~Mironov and A.~Morozov, ``{The Power of Nekrasov Functions},''
  \href{http://dx.doi.org/10.1016/j.physletb.2009.08.061}{{\em Phys.Lett.}
  {\bfseries B680} (2009) 188--194},
\href{http://arxiv.org/abs/0908.2190}{{\ttfamily arXiv:0908.2190 [hep-th]}}.

\bibitem{Mironov:2009by}
A.~Mironov and A.~Morozov, ``{On AGT relation in the case of U(3)},''
  \href{http://dx.doi.org/10.1016/j.nuclphysb.2009.09.011}{{\em Nucl.Phys.}
  {\bfseries B825} (2010) 1--37},
\href{http://arxiv.org/abs/0908.2569}{{\ttfamily arXiv:0908.2569 [hep-th]}}.

\bibitem{Alba:2010qc}
A.~Alba, A.~Fateev, V.~Litvinov, and M.~Tarnopolskiy, ``{On combinatorial
  expansion of the conformal blocks arising from AGT conjecture},''
  \href{http://dx.doi.org/10.1007/s11005-011-0503-z}{{\em Lett.Math.Phys.}
  {\bfseries 98} (2011) 33--64},
\href{http://arxiv.org/abs/1012.1312}{{\ttfamily arXiv:1012.1312 [hep-th]}}.

\bibitem{Belavin:2011js}
A.~Belavin and V.~Belavin, ``{AGT conjecture and Integrable structure of
  Conformal field theory for c=1},''
  \href{http://dx.doi.org/10.1016/j.nuclphysb.2011.04.014}{{\em Nucl.Phys.}
  {\bfseries B850} (2011) 199--213},
\href{http://arxiv.org/abs/1102.0343}{{\ttfamily arXiv:1102.0343 [hep-th]}}.

\bibitem{Marshakov:2009gs}
A.~Marshakov, A.~Mironov, and A.~Morozov, ``{Combinatorial Expansions of
  Conformal Blocks},'' \href{http://dx.doi.org/10.1007/s11232-010-0067-6}{{\em
  Theor.Math.Phys.} {\bfseries 164} (2010) 831--852},
\href{http://arxiv.org/abs/0907.3946}{{\ttfamily arXiv:0907.3946 [hep-th]}}.

\bibitem{Morozov:2013rma}
A.~Morozov and A.~Smirnov, ``{Towards the Proof of AGT Relations with the Help
  of the Generalized Jack Polynomials},''
  \href{http://dx.doi.org/10.1007/s11005-014-0681-6}{{\em Lett.Math.Phys.}
  {\bfseries 104} no.~5, (2014) 585--612},
\href{http://arxiv.org/abs/1307.2576}{{\ttfamily arXiv:1307.2576 [hep-th]}}.

\bibitem{Mironov:2013oaa}
S.~Mironov, An. Morozov, and Y.~Zenkevich, ``{Generalized Jack polynomials and
  the AGT relations for the $SU(3)$ group},''
  \href{http://dx.doi.org/10.1134/S0021364014020076}{{\em JETP Lett.}
  {\bfseries 99} (2014) 109--113},
\href{http://arxiv.org/abs/1312.5732}{{\ttfamily arXiv:1312.5732 [hep-th]}}.

\bibitem{Pestun:2007rz}
V.~Pestun, ``{Localization of gauge theory on a four-sphere and supersymmetric
  Wilson loops},'' \href{http://dx.doi.org/10.1007/s00220-012-1485-0}{{\em
  Commun.Math.Phys.} {\bfseries 313} (2012) 71--129},
\href{http://arxiv.org/abs/0712.2824}{{\ttfamily arXiv:0712.2824 [hep-th]}}.

\bibitem{Nekrasov:2003rj}
N.~Nekrasov and A.~Okounkov, ``{Seiberg-Witten theory and random partitions},''
\href{http://arxiv.org/abs/hep-th/0306238}{{\ttfamily arXiv:hep-th/0306238
  [hep-th]}}.

\bibitem{Nekrasov:2002qd}
N.~Nekrasov, ``{Seiberg-Witten prepotential from instanton counting},''
  \href{http://dx.doi.org/10.4310/ATMP.2003.v7.n5.a4}{{\em
  Adv.Theor.Math.Phys.} {\bfseries 7} (2004) 831--864},
\href{http://arxiv.org/abs/hep-th/0206161}{{\ttfamily arXiv:hep-th/0206161
  [hep-th]}}.

\bibitem{Dotsenko:1984nm}
V.~Dotsenko and V.~Fateev, ``{Conformal Algebra and Multipoint Correlation
  Functions in Two-Dimensional Statistical Models},''
\href{http://dx.doi.org/10.1016/0550-3213(84)90269-4}{{\em Nucl.Phys.}
  {\bfseries B240} (1984) 312}.

\bibitem{Mironov:2010zs}
A.~Mironov, A.~Morozov, and Sh. Shakirov, ``{Conformal blocks as
  Dotsenko-Fateev Integral Discriminants},''
  \href{http://dx.doi.org/10.1142/S0217751X10049141}{{\em Int.J.Mod.Phys.}
  {\bfseries A25} (2010) 3173--3207},
\href{http://arxiv.org/abs/1001.0563}{{\ttfamily arXiv:1001.0563 [hep-th]}}.

\bibitem{Mironov:2009ib}
A.~Mironov, A.~Morozov, and Sh. Shakirov, ``{Matrix Model Conjecture for Exact
  BS Periods and Nekrasov Functions},''
  \href{http://dx.doi.org/10.1007/JHEP02(2010)030}{{\em JHEP} {\bfseries 1002}
  (2010) 030},
\href{http://arxiv.org/abs/0911.5721}{{\ttfamily arXiv:0911.5721 [hep-th]}}.

\bibitem{Mironov:2010xs}
A.~Mironov, A.~Morozov, and Sh. Shakirov, ``{Brezin-Gross-Witten model as 'pure
  gauge' limit of Selberg integrals},''
  \href{http://dx.doi.org/10.1007/JHEP03(2011)102}{{\em JHEP} {\bfseries 1103}
  (2011) 102},
\href{http://arxiv.org/abs/1011.3481}{{\ttfamily arXiv:1011.3481 [hep-th]}}.

\bibitem{Mironov:2010ym}
A.~Mironov, A.~Morozov, and An. Morozov, ``{Conformal blocks and generalized
  Selberg integrals},''
  \href{http://dx.doi.org/10.1016/j.nuclphysb.2010.10.016}{{\em Nucl.Phys.}
  {\bfseries B843} (2011) 534--557},
\href{http://arxiv.org/abs/1003.5752}{{\ttfamily arXiv:1003.5752 [hep-th]}}.

\bibitem{Gamayun:2012ma}
O.~Gamayun, N.~Iorgov, and O.~Lisovyy, ``{Conformal field theory of Painleve
  VI},'' \href{http://dx.doi.org/10.1007/JHEP10(2012)183,
  10.1007/JHEP10(2012)038}{{\em JHEP} {\bfseries 1210} (2012) 038},
\href{http://arxiv.org/abs/1207.0787}{{\ttfamily arXiv:1207.0787 [hep-th]}}.

\bibitem{Litvinov:2013sxa}
A.~Litvinov, S.~Lukyanov, N.~Nekrasov, and A.~Zamolodchikov, ``{Classical
  Conformal Blocks and Painleve VI},''
  \href{http://dx.doi.org/10.1007/JHEP07(2014)144}{{\em JHEP} {\bfseries 1407}
  (2014) 144},
\href{http://arxiv.org/abs/1309.4700}{{\ttfamily arXiv:1309.4700 [hep-th]}}.

\bibitem{Perlmutter:2015iya}
E.~Perlmutter, ``{Virasoro conformal blocks in closed form},''
\href{http://arxiv.org/abs/1502.07742}{{\ttfamily arXiv:1502.07742 [hep-th]}}.

\bibitem{Nemkov2}
N.~Nemkov, ``{On fusion kernel in Liouville theory},''
\href{http://arxiv.org/abs/1409.3537}{{\ttfamily arXiv:1409.3537 [hep-th]}}.

\bibitem{MS}
G.~Moore and N.~Seiberg, ``{Classical and Quantum Conformal Field Theory},''
\href{http://dx.doi.org/10.1007/BF01238857}{{\em Commun.Math.Phys.} {\bfseries
  123} (1989) 177}.

\bibitem{MSlecturesRCFT}
G.~Moore and N.~Seiberg, ``{Lectures on RCFT},'' {\em RU-89-32, YCTP-P13-89,
  C89-08-14} (1989) .

\bibitem{GMMnonpert}
D.~Galakhov, A.~Mironov, and A.~Morozov, ``{S-Duality and Modular
  Transformation as a non-perturbative deformation of the ordinary
  pq-duality},'' \href{http://dx.doi.org/10.1007/JHEP06(2014)050}{{\em JHEP}
  {\bfseries 06} (2014) 050},
\href{http://arxiv.org/abs/1311.7069}{{\ttfamily arXiv:1311.7069 [hep-th]}}.

\bibitem{Iorgov}
N.~Iorgov, O.~Lisovyy, and Yu. Tykhyy, ``{Painleve VI connection problem and
  monodromy of $c=1$ conformal blocks},''
  \href{http://dx.doi.org/10.1007/JHEP12(2013)029}{{\em JHEP} {\bfseries 1312}
  (2013) 029},
\href{http://arxiv.org/abs/1308.4092}{{\ttfamily arXiv:1308.4092 [hep-th]}}.

\bibitem{Verlinde:1988sn}
E.~Verlinde, ``{Fusion Rules and Modular Transformations in 2D Conformal Field
  Theory},''
\href{http://dx.doi.org/10.1016/0550-3213(88)90603-7}{{\em Nucl.Phys.}
  {\bfseries B300} (1988) 360}.

\bibitem{Drukker:2009id}
N.~Drukker, J.~Gomis, T.~Okuda, and J.~Teschner, ``{Gauge Theory Loop Operators
  and Liouville Theory},''
  \href{http://dx.doi.org/10.1007/JHEP02(2010)057}{{\em JHEP} {\bfseries 1002}
  (2010) 057},
\href{http://arxiv.org/abs/0909.1105}{{\ttfamily arXiv:0909.1105 [hep-th]}}.

\bibitem{Alday:2009fs}
L.~Alday, D.~Gaiotto, S.~Gukov, Y.~Tachikawa, and H.~Verlinde, ``{Loop and
  surface operators in N=2 gauge theory and Liouville modular geometry},''
  \href{http://dx.doi.org/10.1007/JHEP01(2010)113}{{\em JHEP} {\bfseries 1001}
  (2010) 113},
\href{http://arxiv.org/abs/0909.0945}{{\ttfamily arXiv:0909.0945 [hep-th]}}.

\end{thebibliography}\endgroup
\end{document}